\hsize=16.6truecm        \hoffset=-0.3truecm      \vsize=24.5truecm
\overfullrule=0pt        \parindent0.7true cm     \voffset0.0truecm
\font\rm=cmr10           \font\bf=cmbx10	  \font\math=cmsy10
\font\it=cmti10                    \tolerance=250
          \font\smtt=cmtt10 at 5pt
\font\brm=cmr10 scaled\magstep1          \font\bit=cmti10 scaled\magstep1
         \font\ssmmath=cmsy10 at 6pt
         \font\ssrm=cmr10 at 6pt
         
\font\biggk=cmmi10 at 12pt               \font\smgk=cmmi10 at 8pt
\font\smmath=cmsy10 at 8pt               \font\sssmmath=cmsy10 at 4pt
\font\bigmath=cmsy10 at 12pt             \font\ssit=cmti10 at 6pt
\font\srm=cmr10 at 8pt   \font\sit=cmti10 at 8pt  \font\sbf=cmbx10 at 8pt
%\font\srm=cmr10 at 9pt   \font\sit=cmti10 at 9pt  \font\sbf=cmbx10 at 9pt
\font\stmath=cmsy10 at 7pt               \font\stgk=cmmi10 at 7pt
\def\cl{\centerline}     \def\ni{\noindent}       
\def\bs{\bigskip}        \def\ms{\medskip}	  
\def\vs#1{\vskip#1pt}    \def\hs#1{\hskip#1pt}    \def\hm#1{\hskip#1em}
\def\vfe{\vfill\eject}   \def\hf{\hfill}          \def\hah{\hf & \hf}
\def\hcr{\hf\cr}            
  \def\PM{\hs1 $\pm$ \hs1} \def\PN{$\pm$ }

\def\smPM{{\hs1\smmath\hs3\char'6 \hs4}} 
  \def\l1{\looseness-1}
\def\msp{\kern.4em}        \def\={= \sk}
\def\kms{km~s$^{-1}$}    \def\kmss{km~s$^{-1}$\kern.25em}  
\def\bigkms{\hbox{km s\raise.8ex\hbox{\rm\hs{0.8}--\hs{0.5}1}}\hs1}
\def\smkms{\hbox{km s\raise.6ex\hbox{\ssrm\hs{0.8}--\hs{0.5}1}}\hs1}
\def\bigkmss{\bigkms\hs3}           
\def\itkms{\hbox{{\it km s}\raise.6ex\hbox{\sit\hs{0.8}--\hs{0.5}1}}\hs1}
\def\sitkms{\hbox{{\sit km s}\raise.6ex\hbox{\ssit\hs{0.8}--\hs{0.5}1}}\hs1}
\def\itamm{\hbox{{\it \AA~mm}\raise.7ex\hbox{\sit\hs{0.8}--\hs{0.5}1}}\hs1}
\def\up#1{\raise.8ex\hbox{\ssrm#1}\hs3}     \def\bsk#1{\baselineskip#1pt}
\def\upnormal#1{\hs{0.5}\raise.8ex\hbox{\srm#1}\hs3}
\def\down#1{\lower.5ex\hbox{\ssrm#1}\hs3}          \def\:{:\hs3}
\def\downit#1{\lower.5ex\hbox{\ssit#1}\hs3}
      \def\sec{\hbox{$''$\hs{-2}.}}  
\def\bigsec{{\raise.8ex\hbox{{\smmath\char'60\char'60}}}\hs{-1}.}
\def\smsec{\hs1\raise1.0ex\hbox{{\sssmmath\char'060\char'060}}.}
\def\ea{{\it \hbox{et al\/}.}}      \def\eas{\ea\hs3}
\def\etal#1{\hbox{{\it et al\/}.\hs{-0.8}$^{#1}$}}  
\def\setal{{\sit et al\/}.,\hs3}

\def\msun{$M\hs{-1.5}_{\odot}$}     \def\msuns{\msun\hs3}
\def\bigmsun{{\bit M}\lower0.5ex\hbox{\math\char'14}} 
\def\smmsun{{\sit M}\lower0.5ex\hbox{\ssmmath\char'14}}
\def\bigmsuns{\bigmsun\hs4}         
\def\rsun{$R_{\odot}$}              \def\rsuns{\rsun\hs3}
\def\bigsim{{\bigmath\char'30}\hs4}

\def\dego{\raise.9ex\hbox{\smmath\char'16}}     
\def\smdego{\raise.9ex\hbox{\ssmmath\char'16}}  
\def\bigdego{\raise.9ex\hbox{\math\char'16}}    
\def\degs{\dego\hs3}   \def\b{\hm{0.5}}  \def\k{\hm{0.3}}   \def\sk{\hm{0.6}}
          
\def\M{\raise.9ex\hbox{\srm m}}      
\def\bigM{\raise.9ex\hbox{\rm m}}     
\def\mag{\raise.8ex\hbox{\srm m}\hs{-1}.}
\def\bigmag{\raise.8ex\hbox{\rm m}\hs{-1}.}
\def\smmag{\raise.8ex\hbox{\ssrm m}\hs{-1}.}
  \def\K2{\hs{-1}{\it K}\hs{-1}$_2$}
\def\smK#1{{\sit K\lower0.5ex\hbox{\ssrm #1}}}
\def\smq{{\sit q}\hs9\srm (= {\sit m}\lower0.5ex\hbox{\ssrm 1}/{\sit m}\lower0.5ex\hbox{\ssrm 2})}
\def\asini{\hbox{$a_1$\hs1sin\hs2$i$}}          
\def\bsini{\hbox{$a_2$\hs2sin\hs2$i$}}          
\def\sma#1sini{{\sit a\/}\lower0.5ex\hbox{\ssrm #1}{\srm\hs1  sin}\hs1{\sit i}}
\def\smasini{{\sit a}\lower0.5ex\hbox{{\ssrm 1}}\hs2 sin\hs2{\sit i} \hs3 (Gm)}
\def\vsini{\hbox{$v$\hs{1.5}sin\hs{1.5}$i$\hs{-0.4}}}  \def\vsinis{\vsini\hs3} 
\def\smfm#1{{\sit f\/}({\sit m\/}\lower0.5ex\hbox{\ssrm #1}) \hs4 ({\sit M}\lower0.5ex\hbox{\ssmmath\char'14})}
\def\smm#1sin3i{{\sit m\/}\lower0.5ex\hbox{\ssrm #1}\srm\hs1sin\raise0.8ex\hbox{\ssrm 3}\hs1{\sit i} \hs3({\sit M}\lower0.5ex\hbox{\ssmmath\char'14})}

          \widowpenalty=1000
\def\zmsin3i{$m_1$\hs1 sin$^3$\hs1 $i$}         \hyphenpenalty=200
\def\m2sin3i{$m_2$\hs1 sin$^3$\hs1 $i$}         \clubpenalty=1000
        \brokenpenalty=1000
\def\cor#1{\hbox{{\it C\hs{-0.8}oravel\/}$^{#1}$}}  
\def\cor{\hbox{$C\hs{-1}oravel\/$}}              \def\cors{\cor\hs4}
\def\smcor{\hbox{\sit C\hs{-0.5}oravel\/}}       \def\smcors{\smcor\hs4}
\def\b-v{\hbox{($B-V$)}\hs4}                     \def\u-b{\hbox{($U-B$)}\hs4}  
\def\smb-v{({\sit B {\smmath\char'0} V\/})}      \def\v{$V$\hs3}
\def\bigb-v{(\hs{-1}{\bit B {\bigmath\char'0}\hs4V\/})\hs4}
\def\smu-b{({\sit U {\smmath\char'0} B\/})}
\def\frac#1/#2{\leavevmode\kern.05em\raise.6ex\hbox{\the\scriptfont0
    #1}\kern-.15em/\kern-.15em\lower.3ex\hbox{\the\scriptfont0 #2}\hs1}
\def\smfrac#1/#2{\leavevmode\kern.05em\raise.6ex\hbox{\the\scriptscriptfont0
    #1}\kern-.15em\raise.1ex\hbox{{\srm /}}\kern-.15em\lower.3ex\hbox{\the\scriptscriptfont0 #2}\hs1}
\def\bigfrac#1/#2{\leavevmode\kern.05em\raise.6ex\hbox{\rm
    #1}\kern-.2em\raise.1ex\hbox{\brm /}\kern-.15em\lower.3ex\hbox{\rm #2}\hs1}
\def\gsim{{\hbox{\raise.5ex\hbox{$>$}}\hskip-.8em\lower.6ex\hbox{$\sim$}}\hs5}
\def\lsim{{\hbox{\raise.4ex\hbox{$<$}}\hskip-.8em\lower.5ex\hbox{$\sim$}}\hs5}
\def\smgsim{{\hbox{\raise.5ex\hbox{{\stgk\char'76}}}\hskip-.7em\lower.5ex\hbox{{\stmath\char'30}}}\hs2}
\def\smlsim{{\hbox{\raise.5ex\hbox{{\stgk\char'74}}}\hskip-.7em\lower.5ex\hbox{{\stmath\char'30}}}\hs2}
\def\biglsim{{\hbox{\raise.5ex\hbox{{\biggk\char'74}}}\hskip-.7em\lower.5ex\hbox{{\bigmath\char'30}}}\hs2}
     
\def\pip{{\hs{-1}\ssmmath\hs2\raise.5ex\hbox{\char'17}}}  
  
\def\section#1{\bs\ni\it#1\ms\rm}
\def\*#1{{\bsk8{\parindent9pt\footnote*{\hs{-5}{\srm #1}}}}}

\def\refs{\vs{20}\cl{\it References}\ms\srm\bsk{12}}
\def\ref#1{\par\ni\parskip=3pt\hangindent=1.66pc\hangafter=1\hskip0.5em(#1)\sk}
\def\r#1{\par\ni\parskip=3pt\hangindent=1.66pc\hangafter=1\hskip0.0em (#1)\sk}
\def\s#1{\par\ni\parskip=3pt\hangindent=2.13pc\hangafter=1 (#1)\sk}

\def\fig#1{\srm\bsk{14.5}\vs{17}\cl{F{\ssrm IG}. #1}\vs2}
\def\aa#1{{\sit A\&A},\k{\sbf #1},\k}     % For refs, 1 digit, hang 1.33 pc. 
   % For refs, 3 digits, hang 2.13 pc.

\def\apj#1{{\sit ApJ},\k{\sbf #1},\k}

\def\ha#1{{\sit HA},\k{\sbf #1},\k}
\def\ibvs#1{{\sit IBVS}, no.~#1,\k}
\def\x#1#2{{\sit #1}, no.~#2,\k}

\def\obs#1{{\sit The Observatory},\k{\sbf #1},\k}
\def\pasp#1{{\sit PASP},\k{\sbf #1},\k}
\def\pdao#1{{\sit PDAO},\k{\sbf #1},\k}

\def\s&t#1{{\sit Sky \& Tel.},\k{\sbf #1},\k}

\def\aq{C. W. Allen, {\sit Astrophysical Quantities} (Athlone, London), 1973}

\def\hd{{\it Henry Draper Catalogue\/}}      \def\hds{\hd\hs3}
\def\hde{{\it Henry Draper Extension\/}}      \def\hdes{\hdes\hs3}
  
\def\HD{\hbox{{\it H\hs{-0.4}D\/}}}          \def\HDs{\HD\hs3}
\def\bigHD{\hbox{{\bit H\hs{-0.4}D\/}}}      \def\bigHDs{\bigHD\hs4}

\def\n2000{I. Ridpath (ed.), {\sit Norton's 2000.0} (Longman, Harlow),
    editions 18 \& 19, 1989 \& 1998}
\def\h{{\sit The Hipparcos and Tycho Catalogues} (ESA SP--1200) (ESA, 
    Noordwijk), 1997}
\def\tycho2{[Announced by] E. H\char'34g \setal \aa{355} L27, 2000.}

\def\hp{{\it Hipparcos\/}}       \def\hps{{\it Hipparcos\/}\hs4}
\def\bighp{{\bit Hipparcos\/}}   \def\bighps{\bighp\hs4}
\def\smhp{{\sit Hipparcos\/}}    \def\smhps{\smhp\hs4}
\def\T2{{\it Tycho 2\/}}         \def\T2s{\T2\hs3}
\def\t{{\it Tycho\/}}            \def\ts{\t\hs3}
\def\smt{{\sit Tycho\/}}         \def\smts{\smt\hs3}
\def\smhp{{\sit Hipparcos\/}}    \def\smhps{\smhp\hs4}
\def\8cat#1{A. H. Batten, J. M. Fletcher \& D. G. MacCarthy, {\sit Eighth 
    Catalogue  of the Orbital Elements of Spectroscopic Binary Systems}\break
    (\pdao{17} #1, 1989).} 
\def\8cat{A. H. Batten, J. M. Fletcher \& D. G. MacCarthy, {\sit Eighth 
    Catalogue  of the Orbital Elements of Spectroscopic Binary Systems}\break
    (\pdao{17} 1989).} 

\def\coll135#1{, in H. A. McAlister \& W. I. Hartkopf (eds.), {\sit 
    Complementary Approaches to Double and Multiple Star Research} ({\sit
    IAU Colloquium}, no.~135) ({\sit ASP Conference Series}, no.~32) (ASP,
    San Francisco), 1992, p.\hs3#1.}  
\def\adams35{W. S. Adams \setal \apj{81} 187, 1935.}

\def\Herschel{I\hs{-2.5}\lower.3ex\hbox{{\smmath\char'76}\hs{-5.2}\lower.3ex
    \hbox{$_{\circ}$}}\hs{-2.3}I\hs4}
\def\H2{I\hs{-1.5}\raise0.3ex\hbox{-\hs{-2.0}-}\hs{-2.0}I\hs{-5.5}\lower.05ex
    \hbox{{\smtt\char'174}\hs{-3.25}\lower.42ex\hbox{$_{\circ}$}}\hs4}

\def\PLO16#1{W. W. Campbell \& J. H. Moore, {\sit Publ.~Lick Obs.}, {\sbf 16}, 
    #1, 1928.}
  
\def\Fig21{A. Baranne, M. Mayor \& J.-L. Poncet, {\sit Vistas in Astr.},
    {\sbf 23}, 279, 1979.  (See Fig.~21 on p.~313.)}
\def\coll88{, in A. G. D. Philip \& D. W. Latham (eds.), {\sit Stellar
       Radial Velocities} ({\sit IAU Coll.}, no.~88) (Davis, Schenectady), 
       1985, p.~}
\def\coll170#1{ in {\sit Precise Stellar Radial Velocities,} eds.~J. B. 
    Hearnshaw \& C. D. Scarfe ({\sit IAU Coll.}, no.~170) ({\sit ASP 
    Conf.~Series}, {\sbf 185}) (ASP, San Francisco), 1999, p.~#1}

\def\hut{P. Hut, \aa{99} 126, 1981.}

\def\title#1{\bsk{16}\rm\topinsert\vskip0pt\endinsert
\cl{SPECTROSCOPIC BINARY ORBITS}\vs2 \cl{FROM PHOTOELECTRIC RADIAL VELOCITIES} 
\vs8\cl{PAPER #1}\vs9\it\cl{By R.F.~Griffin}\vs{-2}\cl{Cambridge Observatories}}

\bsk{17}
\cl{SPECTROSCOPIC BINARY ORBITS}\vs2 \cl{FROM PHOTOELECTRIC RADIAL VELOCITIES} 
\vs{6}\cl{PAPER 171: HD 152028 {\srm AND} HDE 284195}
\cl{(GK D{\srm RACONIS\hs4AND} V1094 T{\srm AURI})}\it\bs
\settabs\+\hskip 1.3 true cm & \hskip 6.0 true cm & 2.0 true cm & \hskip 6.0 
 true cm & \hskip 1.0 true cm & \cr\it
\+ \hah By R. F. Griffin \hah and \hah H. M. J. Boffin\hah \hcr
\+ \hah Cambridge Observatories \hah \hah European Southern Observatory \hah \hcr

\vs{12}\brm\bsk{19.5} The two stars that form the subject of this paper 
are both short-period double-lined eclipsing binaries.  They have
non-circular orbits despite their short periods, both of which are very
close to integral numbers of sidereal days (10.0015 and 9.0127 sidereal days
respectively), making it difficult to achieve uniform phase coverage of the
orbits from a single site in the short term.

Although the \bigHDs type of HD~152028 is G0, the integrated spectral type
of the system must actually be much earlier\: the \bigb-v colour index is
only about \hbox{0\bigmag 37} and the \bighps parallax indicates an
integrated absolute magnitude as bright as +1\bigmag 4.  A published
photometric investigation suggests that the primary star exhibits
\hbox{{\biggk\char'16} Scuti} pulsations, with a period of 0.1138 days.
That period is not present in the radial velocities, but we have identified
a comparable periodicity in the initially excessive residuals
({\biggk\char'33} {\bigmath\char'30} 2 \bigkms ) in the radial velocities of
the primary star: there is an asymmetrical pulsational velocity curve with a
semi-amplitude of \bigsim\hs{-3} 3 \bigkmss and a period of 0.1178 days.
The disagreement between that period and the published photometric one is
\l1 probably because the latter is mistaken.

It is argued from the luminosity of the system, the eclipse light-curves in
{\bit V\/} and \hbox{\bigb-v\hs{-4},} the lengths of the eclipse chords, and
the rotational velocity of the primary star, that the eclipses are
practically central (a total eclipse and a transit) and that the component
stars have the same colour but luminosities that differ by 1\bigmag 0.

HDE~284195 was not observed by \bighp , but its {\bit H\hs{-0.4}DE\/} type
of G0 is in reasonable agreement with its colour and the nature of its
radial-velocity traces.  The rotations of both stars appear to be
pseudo-synchronized to the orbit.

Since this is \lq only' a radial-velocity study and not a comprehensive
discussion of the objects, the orbital inclinations are not formally
determined.  They must be very high, however (otherwise the systems would
not eclipse), and there is specific evidence that the inclination of
HD~152028 is very close to 90\bigdego , so it is permissible to assume that
the factors sin{\raise0.7ex\hbox{{\rm 3}}}\hs2{\bit i\/} in the masses are
close to unity.  In that case the masses are scarcely above the minimum
values, which in the case of HD~152028 are 1.78 and 1.42 \bigmsuns and in
that of HDE~284195 are 1.10 and 1.01 \bigmsun , with uncertainties \biglsim
1\%. 

\vfe\rm\bsk{17.2}\section{Introduction}

This paper presents orbits for two ninth-magnitude binary stars to which
attention has been called by the discovery of eclipses.  Both of them have
proved to be spectroscopically double-lined, so their orbital elements
possess particular significance by giving the masses directly.  Another
characteristic, rather troublesome, that the systems share is the close
approximation of the period of each to an integral number of sidereal days.

HD 152028 is in Draco, about \frac1/4\hs3 of the way from $\zeta$ Draconis
towards $\beta$ Ursae Minoris; it was classified as type G0 in the \hd .  It
had attracted no attention whatsoever from astronomers until \hp$^1$
discovered its eclipses, to which it attributed an incorrect period of
16.960 days; the normal magnitude was found to be 8\mag 76.  The \hps
discovery resulted in the assignment$^2$ of the variable-star designation GK
Draconis.  Very recently Dallaporta \etal3 have published a photometric
investigation that has identified the true period as 9.9742 days and has
demonstrated that there are primary and secondary eclipses of virtually
equal depths, 0\mag 36, but somewhat different widths, 0.050 and 0.039 of
the orbital period.  In addition there is \lq photometric noise' that
becomes less conspicuous during the secondary minimum and is accordingly
attributed to the star that is being eclipsed at that phase; a 2.7-hour
periodicity is identified, and the suggestion is made that the relevant
component of the binary is a $\delta$ Scuti variable.  The \hps parallax is
0\sec 00337 \PN 0\sec 00069, corresponding to a true distance modulus of
about 7\mag 36 \PN 0\mag 5; even without allowance for interstellar
absorption, the absolute magnitude must be as bright as +1\mag 4 with the
same uncertainty of 0\mag 5 --- much brighter than would be expected for a
system with an integrated spectral type of G0 if it were a normal
main-sequence pair.  \hps gives the \b-v colour index as 0\mag 376, while
Dallaporta \etal3 give it as 0\mag 35; it is clearly too blue for a G0 type.

HDE 284195 (BD +21\degs 605) is in Taurus, roughly midway between the
Pleiades and the centre of the Hyades; at ninth magnitude, it was marginally
too faint for inclusion in the original \hds but was picked up in the second
volume of the \hde$^4$, where it was classed G0.  Like HD~152028, it came to
attention only when it was discovered to be an eclipsing system; the discovery
was made by Kaiser$^5$, who initially gave its period as 3.176 days.  Soon
afterwards, Kaiser and five collaborators issued a correction to the period,
explaining$^6$ that the orbit must be non-circular, since the secondary
eclipse did not come at phase .5\: the originally-supposed period of 3.176
days was actually the interval between the secondary eclipse and the ensuing
primary one, the new value for the period being 4.49407 days.  Later still,
Kaiser \& Frey$^7$, acting on information provided privately by L. Marshall
in 1996, further revised the period by doubling it, giving a true value of
8.988476 \PN 0.000007 days.  Meanwhile, Kazarovets \& Samus$^8$ had gazetted
HDE~284195 as a variable star with the designation V1094 Tauri.

Despite all the interest in its photometry, it seems that no actual
magnitudes of HDE~284195 have been published.  The star was not on the \hps
programme, but was observed as a \ts star; the new reductions in the {\it
Tycho 2\/} catalogue$^9$ give \hbox{$V_T$ = 9\mag 052 \PN 0\mag 024,}
\hbox{$B_T$ = 9\mag 833 \PN 0\mag 032}.  According to information to be
found in the \hps introductory volume (ref.~1, {\bf 1}, p.~57) those
magnitudes transform to \hbox{\v = 8\mag 98,} \hbox{\b-v = 0\mag 66,}
evidently with uncertainties of a few hundredths; a recalibration by
Bessell$^{10}$ of the \hps and \ts photometric system would suggest values
of 8\mag 97 and 0\mag 70.

\section{Radial velocities and orbits}

No radial velocities are known to have been published for either star,
although Dallaporta \etal3 mentioned that spectroscopic observations of
HD~152028 were in progress and the advice that Kaiser \& Frey$^7$ ascribed
to Marshall concerning HDE~284195 stemmed from radial-velocity observations
that seem not to have been published.

Both systems have recently been under observation with the \cors
radial-velocity spectrometer at the coud\'e focus of the Cambridge 36-inch
reflector.  Observations of HD~152028 were begun in the 2001 observing
season and promptly demonstrated that the 16.960-day period announced by
\hps was in error.  That is, as a matter of fact, no longer particularly
noteworthy, as it has proved to be the case for quite a number of stars
noted by \hps as eclipsing binaries; this is the third example in this {\it
Magazine}, after HD~44192 (V454~Aur)$^{11}$ and HD~98116 (FO~Leo)$^{12}$,
while others, including HD~175900 (HP~Dra)$^{13}$ and HD~183361
(V2080~Cyg)$^{13}$ have been documented elsewhere.  The period-finding
routine used by the \hps authors, who were of course obliged to deal with
their photometric material on a wholesale basis, seems to have worked
satisfactorily in cases where the variations were quasi-continuous, but did
not cope well with eclipsing systems.  In such cases, most of the
photometric data are clustered at the level of maximum brightness, with
relatively few outliers representing eclipses, and the opportunities for
aliasing --- particularly in the context of the very inappropriate typical
\hps \lq window function' --- were evidently a great deal higher than those
authors appreciated.  It has been pointed out previously (ref.~11, p.~325)
that radial-velocity observations, where the quantity being observed varies
continuously instead of only on rare excursions, offer a much more
favourable basis for establishing periods.  Accordingly it did not take long
to recognize from the radial-velocity data that the true period of HD~152028
is very close to 10 days.  It is only fair to remark here that, although
radial velocities have the advantage over photometry in the initial
discovery of a periodicity, the position is reversed once the approximate
period is known\: whereas radial velocities change gradually over the whole
period, the suddenness of the onset and cessation of eclipses means that the
whole photometric range is spanned in a small fraction of the period and
gives photometry an advantage by a factor typically of 10--100 times in
determining the exact value.

The HD 152028 system has continued under radial-velocity observation up to
the time of writing of this paper, the total number of measurements
available for the determination of the orbit being 50.  They are set out in
Table I.  Although the orbital period is extremely close to the integral
number of 10 sidereal days (= 9.9727 solar days), so close that the object
comes to the meridian at virtually the same set of phases year after year,
it has been possible to improve the phase distribution of the radial
velocities by observing at different hour angles.  That is facilitated by
the high declination (+68\dego ) of the object; HD~152028 is therefore
circumpolar as seen from Cambridge (and indeed from latitudes all the way
down to the Tropic of Cancer), and so can in principle be observed at {\it
any\/} hour angle.  In actual practice, however, it can not be observed with
the Cambridge system at extreme hour angles because the coud\'e beam of the
36-inch reflector travels {\it up\/} the polar axis and the \cors occupies
the space to the north of the telescope, obstructing the view of the sky
below the North Celestial Pole.

In almost all observations the velocities of both components have been
found, usually but not invariably from a single trace that includes both \lq
dips', but the dip given by the secondary is extremely weak and often
challenges the limits of what one can attempt to measure with the \cor ,
{\it cf}.~Fig~1.  Even the dip that represents the primary is weak.  It is
obvious not only from the character of the radial-velocity traces but also
from the absolute magnitude and from the colour index of the system
(remarked upon in the {\it Introduction\/} above) that the integrated
spectral type must be a great deal earlier than the G0 given in the \hd ;
that is borne out by the finding of Dallaporta \etal3 that the primary is
itself an intrinsic variable of the $\delta$ Scuti type, implying a spectral
class of late A or early F.

This is the point at which it is convenient for the writers to explain the
origin of their collaboration.  A~paper only {\it somewhat\/} analogous to
this one was drafted by the first-named author, submitted, and refereed by
the second author.  In the form in which it was initially submitted for
publication, the r.m.s.~residual of the velocities of the primary star of
HD~152028 was as large as 2.0 \kmss --- much worse than might have been
expected by comparison, for example, with the sub-\kmss residuals for even
the secondary component of HDE~284195.  The residuals were not related to
the published$^3$ $\delta$ Scuti periodicity of 0.1137601 days.  In~the
course of refereeing the paper, H. M. J. B. identified in the residuals a
somewhat different periodicity of 8.49 cycles per day (period 0.1178 days).
That discovery led to a collaboration that has resulted in a paper that is
much more substantial and informative than the original one.  

We have been able to apply to the radial-velocity data a program intended
for the solution of orbits of triple stellar systems, to solve
simultaneously for the elements of the binary orbit and for the pulsation of
the primary star.  In~doing so, we have necessarily treated the pulsational
velocity curve as if it could be represented by a Keplerian orbit, but
within the uncertainties of its determination that has proved to be a fully
acceptable approximation.  Much of the excess \lq noise' that at first
existed in the radial velocities of the primary star has disappeared as a
result of the identification of the 0.1178-day period; the remaining
r.m.s.~residual is only 0.70 \kms .  Even so, there is evidence (the
presentation of which is deferred to the {\it Discussion\/} section below)
that the pulsation is not strictly repetitive, so it is likely that its
complete characterization would allow a modest further reduction in the
primary star's radial-velocity residuals.

Here, the principal concern is with the primary--secondary orbit, which is
plotted in Fig.~2.  The~velocities that are plotted there for the primary
star are not the directly observed values but have been corrected for the
respective computed pulsational contributions.  Those contributions are
explicitly listed in the seventh column of Table I; the previous column
tabulates the pulsational phase.  To obtain near-equality of the weighted
variances of the velocities of the two components, it has been necessary to
attribute to the secondary's velocities a weight of only \frac1/{25} .
Table II presents the orbital elements of the binary system, and the
characteristics of the short-period oscillation of the primary star in terms
\l1 of the equivalent Keplerian elements.

\bs\cl{T{\srm ABLE} II}\cl{{\it Orbital and pulsational elements for HD
152028 \hs4{\rm (}\hs{-1}GK Draconis{\rm )}}}
\vs9{\srm\bsk{10}
\halign{\hm{12.5}#\hfil\hm1&\hfil#&#\hfil\hm4&\hfil#&#\hfil\cr
{\hm1\sit Element}&{\sit Orbit}\span&\hs{-3}{\sit Pulsation}\span\cr
\noalign{\vs5}
{\sit P} \hs{3.6} (days)&9.97380&\smPM 0.00019&0.1177753&\smPM 0.0000005\cr
{\sit T} \hs{3.3} (MJD) &52558.36&\smPM 0.05&52558.225&\smPM 0.004\cr
{\smgk\char'15} \hs5 (\smkms )&+2.23&\smPM 0.11\cr
\smK1 \hs3 (\smkms )&64.97&\smPM 0.13&2.62&\smPM 0.17\cr
\smK2 \hs3 (\smkms )&81.14&\smPM 0.6\cr
{\sit q}&1.249&\smPM 0.010\cr
{\sit e}&0.0815&\smPM 0.0021&0.26&\smPM 0.06\cr
\hs{0.5}{\smgk\char'041} \hs3 (degrees)&82.0&\smPM 1.7&96&\smPM 13\cr
\noalign{\vs5}
\sma1sini \hs3 (Gm)&8.881&\smPM 0.018&0.0041&\smPM 0.0003\cr
\sma2sini \hs3 (Gm)&11.09&\smPM 0.09\cr
%{\sit f\/}({\sit m\/}) \hs3 ({\sit M}\lower0.5ex\hbox{\ssmmath\char'14})
\smfm1&0.281&\smPM 0.002\cr
\smfm2&0.548&\smPM 0.013\cr
\smm1sin3i&1.776&\smPM 0.021\cr
\smm2sin3i&1.422&\smPM 0.013\cr
\noalign{\vs5}
R.m.s.\hs2 residual \hs2 (\smkms )&&\hm{-2} 0.70 (Primary)\cr
&&\hm{-2} 3.5 (Secondary)\cr}}\rm\bs

Although slightly fainter and of nominally the same spectral type,
HDE~284195 is an altogether easier object to observe (and to discuss!) than
HD~152028.  Fig.~3 gives an example of a radial-velocity trace, showing two
dips that, while not of very generous depth, readily give velocities whose
characteristic accuracy is better than 1 \kms .  The Cambridge observations
did not start until 2002 September, so all of the 44 measurements that are
listed in Table III are from the 2002/3 observing season.  Again the orbital
period proves to be very nearly an integral number of sidereal days (in this
case 9 days), but the coincidence is not as exact as in the case of
HD~152028: the period is longer by about eighteen minutes, so in each
successive calendar month any given phase is presented at an hour angle that
is progressively displaced about an hour westwards.  That has perversely
meant that towards the end of the observing season, when observations were
necessarily made at westerly hour angles, they fell at much the same set of
phases as those made earlier in the season when the free choice of hour
angle usually led to their being made near the meridian!  Thus the orbit
plot (Fig.~4) shows conspicuously bunched data points, although the
determination of the orbit is scarcely impaired by the bunching.  Eight of
the measurements have not been useable in the solution of the orbit owing to
their being taken when the object was single-lined and the velocities of the
two components were too similar to one another to be separately
determinable; in those cases the observations were reduced as if they were
truly single and the result plotted with an open symbol in Fig.~4.  The 36
remaining observations showed HDE~284195 as double-lined or at least (in the
case of blends) could be reduced as such; the velocities given by the weaker
secondary dip were less accurately measureable than those of the primary,
and it was found necessary to attribute a weight of \frac1/3\hs3 to them to
bring the variances for the two components into approximate equality.  On
that basis the orbital elements are:

\vs{15}
\settabs\+\hskip 2.0 true cm & \hskip 0.6 true cm & \hskip 6.0 true cm & \hskip
 1.5 true cm & \hskip 6.0 true cm \cr 
\+ & $P$ & \= 8.9881 \PM 0.0005 days & ($T$)$_{13}$ & \= MJD 52656.260 \PM 0.007\cr
\+ & $\gamma$ & \= +4.59 \PM 0.07 \kms & \asini & \= 7.772 \PM 0.015 Gm \cr
\+ & \hs{-1}$K_1$ & \= 65.30 \PM 0.12 \kms & \bsini & \= 8.448 \PM 0.0.024 Gm \cr
\+ & \hs{-1}$K_2$ & \= 70.98 \PM 0.20 \kms & $f(m_1)$ & \= 0.2321 \PM 0.0013 \msun \cr
\+ & $q$ & \= 1.087 \PM 0.004 (= $m_1/m_2$) & $f(m_2)$ & \= 0.2980 \PM 0.0026 
\msun \cr
\+ & $e$ & \= 0.2697 \PM 0.0018& \zmsin3i & \= 1.099 \PM 0.007 \msun \cr
\+ & $\omega$ & \= 333.2 \PM 0.3 degrees & \m2sin3i & \= 1.011 \PM 0.005 \msun
\cr\vs3
\cl{R.m.s. residual = 0.46 \kmss (primary), 0.83 \kmss (secondary)\hm2}
\vs{15}

The radial velocities of the two stars that constitute HD~284195 are
computed to differ by nearly 33~\kmss at the conjunctions, so it would be
quite possible (and interesting) to watch the signature of the eclipsing
component grow at the expense of that of the eclipsed one at such times, as
was in fact done in the case (described in Paper 160$^{11}$) of HD~44192.
In the present case, however, the object was inaccessible to observations at
the relevant phases; moreover, it has been concluded$^{14}$, after a sober
assessment of the scientific value of such observations, that they do little
more than serve as an inefficient substitute for photometry, so no excuse is
really warranted for their absence.  Primary eclipses (those in which the
secondary star obscures the primary, at phase .241) will be favourably
presented for observers near the longitude (0\dego ) of Cambridge throughout
the next observing season, every ninth night from 2003 September~1.13 to
2004~March~16.86.

In the case of HD~152028 the components' radial velocities differ by
scarcely more than 2 \kmss at the conjunctions, so there would be no
particular interest in making radial-velocity observations at eclipses: the
dip would appear single-lined but the proportions of it attributable to the
respective components would vary.  One can determine retrospectively that
the primary eclipse, which is centred at phase .526, must have been in
progress at the time of the single-lined observation of 2001~June~25, and
that egress from eclipse had probably not quite finished at the times of the
observations of 2001~November~1, 2002~September~26, and perhaps
2002~October~6.

\section{Discussion --- HD 152028}

In Fig.~5 we present the \hps photometry folded on the true orbital period,
superseding the diagram in ref.~1, vol.~{\bf 12}.  Following {\it
photometric\/} convention, the zero-point of the abscissae in the Figure is
taken as the phase of one of the conjunctions in the spectroscopic orbit
(the one at which the eclipse deemed by Dallaporta \etal3 to be the primary
one takes place).  In order to get the phasing of the eclipses observed by
\hps to agree with those of the conjunctions in the recently observed orbit,
it has been found necessary to refine the period from the 9.97380 \PN 0.00019
days determined above to 9.9742 days; the latter value is exactly the
one found photometrically by Dallaporta \eas (9.9742 \PN 0.0001).  The
change, though small in absolute terms, is somewhat uncomfortable in relation
to the formal standard deviation of the spectroscopically determined period.
The eclipse light-curves here look more plausible than those in the \hps
reference, where one of them is a chain of points in practically a vertical
line.  9.9742 solar days equal 10.0015 sidereal days, so from the point of
view of an observer at a fixed longitude it takes 1000 days for the phasing
of observing opportunities to change by 0.15 days; opportunities would not
average out uniformly until a slippage of a complete day had occurred,
implying a lapse of getting on for 7000 days or say 18 years.

All the evidence points towards a spectral type much earlier than the \HDs
class of G0 for HD~152028.  There is the colour index of$^{1,3}$ about 0\mag
37, the trigonometrically-derived$^1$ absolute magnitude of about +1\mag 4,
the minimum masses of 1.78 and 1.42 \msuns found from the orbital elements
above, and finally the assignment$^3$ of part of the photometric variations
to $\delta$ Scuti pulsations, which do not occur in G-type stars.  On the
basis of actual masses that are scarcely greater than the minimum ones,
which must be the case since the fact that the system exhibits eclipses
demonstrates that sin\hs2$i \sim 1$, one could expect main-sequence types of
about A9 and F3.  Confusingly, however, Dallaporta \etal3 found that there
is no perceptible change of \b-v colour index during either of the
eclipses\: the implication is that the components have identical colour
indices despite their very unequal masses.

It is the primary that Dallaporta \eas found to be the pulsating star, since
the \lq photometric noise' that they attributed to pulsations was
considerably reduced when that star was in eclipse.  The light-curves at
both eclipses appear to have rather flat minima, as if they represented a
total eclipse and a transit, but in both cases the relatively flat part at
the bottom of the minimum is not truly horizontal but slopes down towards
egress, so the flat-bottomed appearance {\it may\/} be a misleading effect
of noise.  Since the components seem to have identical colour indices they
can be supposed also to have identical surface brightnesses.  In that case
the 0\mag 36 depths of both eclipses tell us that at each eclipse there is a
loss of 28\% of the total of the stars' visible surface areas.  That would
agree with the idea of the eclipses being total, with the stars having
relative surface areas, and therefore luminosities, in the ratio 72 to 28,
very close to one stellar magnitude.  Their actual absolute $V$ magnitudes 
would need to be +1\mag 8 and + 2\mag 8 to constitute the +1\mag 4 found by
\hps for the system as a whole.

The relative transverse velocities of the component stars at the times of
the eclipses are computed to be 134 \kmss at the conjunction where the
primary is eclipsed (on the apastron side of the orbit --- the eclipse
called by Dallaporta \etal3 ``Min.~I'') and 157 \kmss at the other.
Dallaporta \eas say that those eclipses last for about 0.050 and 0.039 of
the orbital period, {\it i.e}.~0.50 and 0.39 days, respectively.  Simply by
multiplying together the durations and velocities, we obtain the lengths of
the eclipse chords\: they are respectively 5.8 and 5.3 Gm.  It is physically
impossible for the length of the eclipse chord on the apastron side of the
orbit to be longer than that at the eclipse near periastron, since any
departure of the orbital inclination from 90\degs must result in the
apastron chord being {\it shorter\/} than the other.  Dallaporta \ea 's
graphs of the eclipse photometry demonstrate that photometric noise
seriously limits the accuracy with which the durations of the eclipses can be
determined, so we probably do not need to worry about the apparent
discrepancy, and for the purposes of the present discussion we adopt a mean
eclipse chord of 5.55 Gm or 8.0 \rsuns for both eclipses.  Even if the
eclipses are central (orbital inclination 90\degs exactly), that means that
the sum of the radii of the stars is 4.0 \rsun , at first sight a
surprisingly large value that goes a long way to reinforce the idea that the
eclipses {\it are\/} practically central.

On the basis (proposed above) that the stars have equal surface brightnesses
and that the areas of their surfaces are in the ratio 72 to 28 or 2.57 to 1,
their individual radii must be approximately 2.5 \rsuns for the primary and
1.5 \rsuns for the secondary.  The latter radius is 10--20\% larger than the
tabular$^{15}$ value for a main-sequence star of the colour index of the
HD~152028 system (F3\hs2V, 0\mag 37), but is actually {\it near the lower
bound\/} of the distribution of radii with colour index in the graph shown
by Andersen$^{16}$ of values well determined from eclipsing binaries, so we
can view it with some equanimity.  Neither the radius nor the colour of the
primary star is appropriate to a main-sequence star of its mass, but we
might best regard that star not as an anomalous object but as a somewhat
evolved one, having begun its evolution towards the giant branch in the H--R
diagram by moving towards the right from an initial position corresponding
to a main-sequence type of about A9.  Its present type might be estimated at
F2\hs2III--IV.

We turn now to the evidence for pulsational instability of the primary star.
Dallaporta \etal3 were able to reduce the initially apparent
\lq photometric noise' by demonstrating in it a periodicity, which they
attributed to $\delta$~Scuti variability, of a fraction of a day.
Similarly, through the identification of a quite similar period in the
radial velocities, a major reduction has been effected in what initially
appeared to be \lq radial-velocity noise'.

There is an obvious embarrassment over the question of the pulsational
period because, whereas the photometrists have identified a period of
0.1136701 days (curiously enough stated to seven significant figures but
attributed an uncertainty of as much as 0.0003 days), the period that suits
the radial velocities is 0.1177753 days with an uncertainty of only half a
millionth of a day or some 40 milliseconds.  We attribute the discrepancy to
an error in the photometric period, which was established from measurements
that were all taken within the same small range of {\it orbital\/} phases
and therefore at epochs differing by multiples of 9.974 days.  During one
orbital period the number of pulsational periods according to the
photometric value is 87.677, while according to the radial-velocity value it
is 84.676.  It may well be supposed that the photometric period could not be
determined within one night to a precision sufficient to permit an
unambiguous cycle count to be made to an epoch 10 days away (the actual
intervals between the four nights utilized by Dallaporta \etal3 in their
pulsational analysis were 40, 50, and 10 days), and they made an error of
three cycles per orbital period.  If they had not seemingly arbitrarily
restricted themselves to photometry made on one particular night of the
orbital cycle they would have picked up the mistake.  We can demonstrate the
merit of the radial-velocity period photometrically from the \hps data: in
Fig.~6 they are plotted on the photometric period and in Fig.~7 on the
radial-velocity one.  Only in Fig.~7 do the data show a systematic run with
phase.  It should be noted that the phasing of the photometric wave
demonstrated in Fig.~7 is not secure: in the $\sim$4000 days or 35,000
photometric periods since the \hps epoch the 1-$\sigma$ uncertainty in the
period multiplies up to an interval of 0.017 days or 0.14 pulsational
periods.

The short-period radial-velocity variation was established from observations
that were made on different nights and therefore in cycles far removed from
one another.  The observed variation must be slightly blurred because the
radial-velocity integrations typically lasted nearly half an hour or 0.02
days, about one-sixth of the pulsational period.  There seemed to be
possible merit in observing the pulsation in \lq real time', and by ignoring
the secondary component of the binary system the integration times could be
somewhat curtailed without loss of precision in the velocities of the
primary.  To that intent HD~152028 was observed continuously in a dedicated
four-hour interval on the night of 2003 April 17/18.  Only the region of
velocity space that included the primary dip was scanned, in a series of
fifteen 960-second integrations starting at 1017-second intervals, the
latter being one-tenth of the period.  The results are listed in Table IV.

\bs\rm\cl{T{\srm ABLE} IV}\vs{-3}
\cl{\it Additional radial-velocity observations of HD 152028 A}\vs9
{\srm  \bsk{10}  \tabskip2em
\halign{\hm2\hf#\hm{-1.5}&\hf#&\hf#\cr
\hm{2.6}\sit Hel.~Date {\srm (}\hs{-1}UT{\srm )}\hm{-3}&&\sit Velocity\hm{-0.2}\cr
&&\sitkms\hm{-0.4}\cr
\noalign{\vs5}
2003 April&17.988&+51.7\cr
&18.000&+52.3\cr
&18.011&+50.5\cr
&18.023&+47.7\cr
&18.035&+44.9\cr}
\vs{-72}\moveright5.2truecm\vbox{
\halign{\hm2\hf#\hm{-1.5}&\hf#&\hf#\cr
\hm{2.6}\sit Hel.~Date {\srm (}\hs{-1}UT{\srm )}\hm{-3}&&\sit Velocity\hm{-0.2}\cr
&&\sitkms\hm{-0.4}\cr
\noalign{\vs5}
2003 April&18.047&+43.4\cr
&18.058&+42.5\cr
&18.070&+45.1\cr
&18.082&+46.3\cr
&18.094&+47.4\cr}
\vs{-72}\moveright5.2truecm\vbox{
\halign{\hm2\hf#\hm{-1.5}&\hf#&\hf#\cr
\hm{2.6}\sit Hel.~Date {\srm (}\hs{-1}UT{\srm )}\hm{-3}&&\sit Velocity\hm{-0.2}\cr
&&\sitkms\hm{-0.4}\cr
\noalign{\vs5}
2003 April&18.105&+48.9\cr
&18.117&+49.0\cr
&18.129&+47.2\cr
&18.141&+44.2\cr
&18.152&+40.7\cr}}}}

\bs In Figs.~8 and 9 the observed radial velocities are plotted against
pulsational phase after the large contribution from the computed orbital
velocity variation has been removed; the points corresponding to the
observations listed in Table I and in Table IV are distinguished from one
another by the use of different plotting symbols.  In Fig.~8 the velocity
curve is that derived from the observations in Table I alone and corresponds
to the pulsational elements given in Table II.  It is seen that the Table IV
velocities do not fit it very well: they suggest a curve of considerably
larger amplitude and possibly slightly shifted phase.  The difference in
amplitudes is much greater than could be explained by the somewhat sharper
time-resolution of the new measurements.  The curve in Fig.~9 is drawn to
suit principally the Table IV points, although the others had to be retained
in the solution at some level (they were given a weight of \frac1/{50}),
because they are necessary for the derivation of the orbital velocity curve
which must be known before the pulsational contribution to the velocities
can be isolated.  The r.m.s.~residual of the Table IV points from the curve
drawn to fit them is only 0.41 \kms , but the curve definitely does not suit
the other points as well as the one plotted in Fig.~8.  (There is not a
one-to-one correspondence between the positions of the points plotted in the
two diagrams, owing to the re-computation of the {\it orbital\/} velocity
variation with the greatly changed weighting system of Fig.~9.)

This exercise tends to show that the pulsation responsible for the
photometric and radial-velocity oscillations, while maintaining clockwork
regularity in phase over an interval of years, does not maintain its form
very exactly.  A tentative conclusion, suggested by analogy with other
examples of pulsating stars, might be that there are additional
periodicities with periods rather similar to the principal one, producing a
beat phenomenon that is manifested in the apparent discrepancy between the
velocity curves seen in Figs.~8 and 9.  It seems fairest to adopt --- as has
been done in Table II and is illustrated in Fig~8 --- the curve that is
assembled from observations made in random and well separated cycles, as
being more representative of the general behaviour of the star than the curve
determined at a single epoch from the observations listed in Table IV and
plotted in Fig.~9.

It is possible to use the relationship between period, luminosity, mass, and
temperature ($e.g$.~ref.~17, Equation 2) to obtain the bolometric magnitude
of HD~152028~A as a function of the pulsation constant $Q$.  We derive:\par
\cl{$M_V$ = 6.78 + 3.33 log $Q$,} \ni where we have assumed a bolometric 
correction of 0\mag 11.  All the quantities that needed to be entered into
the period--luminosity relationship to obtain the above result are very well
constrained.  If the period of 0.117775 days that we have detected
represents the fundamental radial mode, for which the corresponding value of
$Q$ is$^{17}$ 0.033 days, we obtain $M_V$ = +1\mag85, in full agreement with
the value obtained above from the \hps photometry and the eclipse depth.

The difficulty of measuring the radial velocity of the secondary star has
seemed not to be the same on different occasions, as if the secondary too
were of varying character, and certainly many of the residuals from the
computed velocity curve are appalling and without precedent in this work.
There is no clear periodicity in the residuals; although a period analysis
seems to indicate a possible period close to 0.4 days we would not give much
for its reality.  If the seeming variations and residuals stem simply from
observational difficulties and errors resulting from the marginal
observability of that star they are regretted, but the possibility is not
absolutely ruled out.  We note that the position of the star in the H--R
diagram is within the area populated by $\gamma$ Doradus variables, which
have periods in the range 0.4--3 days.

In the absence of adequate experience concerning stars such as those
involved in HD~152028, no reliable deductions can be made from the observed
strengths of the dips seen in radial-velocity traces.  The widths of those
dips may nevertheless be significant.  The widths found from radial-velocity
traces of the secondary star are so uncertain and scatter so widely that it
would be unwise to try to make any deduction from them.  The mean value for
the primary, however, is quite accurately established; if interpreted purely
as the effect of stellar rotation, it yields a \vsinis of 14 \kms .  In view
of the very high inclination that we know to characterize the orbit and that
can reasonably be expected to apply also to the rotational axis, the same
figure could be taken to represent the actual equatorial velocity.  If we
further assume the rotation to be pseudo-synchronized$^{18}$ with the
orbital revolution, the rotation period is 9.6 days and a stellar radius of
2.6 \rsuns is derived.  It could be argued that the stellar pulsation may
broaden the line profile and falsify the \vsinis value derived from it, so
2.6~\rsuns must represent a maximum value.  The pulsational amplitude is,
however, small compared with the apparent rotational velocity, so its
contribution to the line-width is probably minimal.  Moreover, the
similarity of the derived radius to the one found from the length of the
eclipse chord --- a minimum value that seems difficult to refute ---
encourages a belief that the various considerations are converging towards a
realistic model of the system, one that harmonizes, moreover, with the
trigonometrical parallax.

\vfe\section{Discussion --- HDE 284195}

HDE 284195 seems to be a much more straightforward double-lined binary
system than HD~152028.  The disparity in equivalent widths of the two dips
seen in radial-velocity traces ({\it cf}.~Fig.~3) suggests a difference in
\v luminosity of 0\mag 5--0\mag 6, corresponding to about three spectral
sub-types; the mass difference of about 9\% agrees.  The masses of 1.10 and
1.01 \msuns given by the orbit above on the assumption (validated by the
eclipses) that sin\hs2$i \sim 1$ would suggest the types to be F9\hs2V and
G2\hs2V; the integrated colour index of 0\mag 66 (or according to the
Bessell calibration$^{10}$ 0\mag 70) derived from the {\it Tycho 2\/}
photometry might urge us to adopt slightly later types, such as G0 + G3 or
even G1 + G4, but it is to be recalled that the standard error of that index
is as much as 0\mag 04.

As in the case of HD 152028, the observed projected rotational velocities
\vsinis for the components of HDE~284195 may be considered as the actual 
equatorial velocities.  The mean values are 9 and 7 \kmss for the primary
and secondary, respectively, with formal standard deviations that are
smaller than the lower limit of 1 \kmss that is ever claimed for rotational
velocities determined from radial-velocity traces.  If the radii of the
stars are estimated at 1.1 and 1.0 \rsun , respectively, the periods of
their rotations are found to be 6.2 and 7.2 days.  At the orbital
eccentricity of 0.27 found for HDE~284195, the pseudo-synchronous rotational
period$^{18}$ is shorter than the orbital period by a factor of 1.49, making
it 6.0 days, so we can be pretty certain that both stars are
pseudo-synchronized.  

\vs{15}\sit\ni Note added in proof\srm

\bsk{14} A preprint of a paper to be published in {\sit A\&A\/} by Zwitter
{\sit et al}.~(a consortium whose membership overlaps that of Dallaporta
\hbox{{\sit et al}.\up3\hs{-3}}), referring in part to HD~152028,~has just 
appeared.  It gives radial velocities that seem to be more scattered than
those presented in the present paper and does not throw any light on the
pulsational period.  In fact a light-curve analysis is attempted on the
basis of a plot (exactly like Fig.~5 here) of the raw \smhps and \smts
photometry, by way of demonstrating what might be obtainable from such data
rather than in an effort to obtain the best possible values for the stellar
parameters.

\refs\ref1 \h , {\sbf 8}, p.~1648; {\sbf 11}, p.~P19; {\sbf 12}, p.~A360.  
\ref2  E. V. Kazarovets \setal \ibvs{4659} 1999.  
\ref3  S. Dallaporta \setal \ibvs{5312} 2002.  
\ref4  A. J. Cannon \& M. W. Mayall, \ha{112} 55, 1949.  
\ref5  D. H. Kaiser, \ibvs{4119} 1994.  
\ref6  D. H. Kaiser \setal \ibvs{4168} 1995.  
\ref7  D. H. Kaiser \& G. Frey, \ibvs{4544} 1998.  
\ref8  E. V. Kazarovets \& N. N. Samus, \ibvs{4471} 1997.
\ref9  \tycho2 
\r{10} M. S. Bessell, \pasp{112} 961, 2000.
\r{11} R. F. Griffin, \obs{121} 315, 2001 (Paper 160).  
\r{12} R. F. Griffin, \obs{122} 355, 2002.  
\r{13} M. Kupri\'nska-Winiarska \setal \ibvs{4823} 2002.  
\r{14} R. F. Griffin, \obs{123} 129, 2003 (Paper 170).
\r{15} \aq , p.~209.
\r{16} J. Andersen, {\sit A\&A Review}, {\sbf 3}, 91, 1991 (see Fig.~3 on
       p.~105). 
\r{17} M. Breger, in M. Breger \& M. H. Montgomery (eds.), {\sit Delta Scuti 
       and Related Stars} ({\sit ASP Conf.~Series}, {\sbf 210}), 2001, p.~3.
\r{18} \hut 

\vfe
\fig1 Radial-velocity trace of HD 152028 (GK Draconis),
obtained with the Cambridge \smcors on 2003 March 1 and showing the unequal
dips.

\fig2  The observed radial velocities of HD 152028 (GK Draconis) plotted as a 
function of phase, with the velocity curves corresponding to the adopted
orbital elements drawn through them.  Squares represent the velocities of
the primary, circles those of the secondary.  The open diamonds indicate
\lq single-lined' measurements, where the velocities of the components were
so similar to one another that they could not be assigned separately; they
were not used in the solution of the orbit.  The measurements of the primary
star have been corrected for the pulsational radial-velocity variations on
the basis of the averaged pulsational \lq elements' listed in Table II.

\fig3  \cl{Radial-velocity trace of HDE 284195 (V1094 Tauri), obtained with 
the Cambridge \smcors on 2003 February 18.}

\fig4  The observed radial velocities of HDE 284195 (V1094 Tauri) plotted as 
a function of phase, with the velocity curves corresponding to the adopted
orbital elements drawn through them.  Squares represent the velocities of
the primary, circles those of the secondary.  The open diamonds indicate
\lq single-lined' measurements, which were not taken into account in the
solution of the orbit.

\fig5  \cl{The \smhps photometry of HD~152028, phased to the true orbital 
period of 9.9742 days.}

\fig6  \cl{The \smhps photometry of HD~152028, phased to the pulsational 
period given by Dallaporta {\sit et al}.\up3\hs{-3}.}

\fig7  The \smhps photometry of HD~152028, phased to the pulsational period
found in the radial velocities of the primary star.  The fact that the
photometry shows a clear variation with phase in this diagram (and not in
Fig.~6) provides convincing evidence that {\sit this\/} period is the
correct one.  The outlying low points (there are others that fall below the
bottom of the diagram) are explained by the eclipses in the system.  There
would be obvious advantage in \lq cleaning' the photometry of the
pulsational variation documented by this Figure before using it in any
discussion of the eclipse light-curves.

\fig8  The pulsational radial-velocity curve of the primary star in HD
152028, established from the observations given in Table I (plotted here as
filled squares) by a simultaneous solution of the orbital and pulsational
variations.  The \lq elements' of the latter, specified as if they were
Keplerian orbital elements, are included in Table II.  The open star symbols
represent radial-velocity observations made in a continuous run covering
1\hs{-1}\smfrac1/2\hs2 cycles of the pulsation on the night of 2003 April
17/18.  They were not utilized at all in the solution of the orbital and
pulsational elements; they are seen to confirm the variation in a general
way, but to exhibit a distinctly greater amplitude.

\fig9  As Fig.~8, but this time the starred observations were given by far
the greatest weight in the calculation of the orbital and pulsational
elements.  To assist the eye to appreciate which points the computed
velocity curve is trying to match, the filled or open characters of the
plotted points have been reversed in comparison with Fig.~8.  The
larger-than-average amplitude of the pulsation on the night when the \lq
star' points were observed is obvious.

\vfe
\nopagenumbers
\hsize=11truecm          \vsize=24.0 true cm     
\hoffset=2.9truecm       \voffset0.6truecm
\baselineskip=12pt	 \overfullrule=0pt        \parindent0truecm
\font\rm=cmr10           \font\bf=cmbx10	  \font\math=cmsy10
\font\it=cmti10                    \tolerance=250
\font\brm=cmr10 scaled\magstep1          \font\bit=cmti10 scaled\magstep1
                  \font\ssrm=cmr10 at 6pt
\font\srm=cmr10 at 8pt   \font\sit=cmti10 at 8pt  \font\sbf=cmbx10 at 8pt
\font\smdag=cmsy10 at 8pt\font\ssit=cmti10 at 6pt 
\def\cl{\centerline}     \def\ni{\noindent}       
\def\bs{\bigskip}        \def\ms{\medskip}	  
\def\vs#1{\vskip#1pt}         \parskip=0pt
\def\vfe{\vfill\eject}   \def\hf{\hfil}          \def\hah{\hfill & \hfill}
\def\hcr{\hfill & \cr}    
  \def\PM{\hs1 $\pm$ \hs1}

\def\up#1{\raise.8ex\hbox{\ssrm#1}}     \def\bsk#1{\baselineskip#1pt}

\def\sb{\rlap{\hs{0.4}\raise.7ex\hbox{\smdag\char'171}}}
\def\sc{\rlap{\hs{0.4}\raise.7ex\hbox{\smdag\char'172}}}
\def\sc{\rlap{\hs{0.4}\raise.7ex\hbox{\smdag\char'172}}}
\def\sd{\rlap{\hs{0.4}\raise.7ex\hbox{\smdag\char'170}}}
\def\se{\rlap{\hs{0.4}\raise.7ex\hbox{\smdag\char'173}}}
\def\sf{\rlap{\hs{0.4}\raise.7ex\hbox{\smdag\char'153}}}

\def\si{\rlap{\hs{0.4}\raise.7ex\hbox{\smdag\char'172\char'172}}}
\def\s#1{\rlap{\hs{0.4}\raise.7ex\hbox{\ssrm#1}}}
\def\sC{\rlap{\hs{0.4}\raise.7ex\hbox{\ssrm C}}}
\def\sI{\rlap{\hs{0.4}\raise.7ex\hbox{\ssrm I}}}
\def\sN{\rlap{\hs{0.4}\raise.7ex\hbox{\ssrm N}}}
\def\sT{\rlap{\hs{0.4}\raise.7ex\hbox{\ssrm T}}}
\def\sU{\rlap{\hs{0.4}\raise.7ex\hbox{\ssrm U}}}
\def\sW{\rlap{\hs{0.4}\raise.7ex\hbox{\ssrm W}}}
       
\def\tb{\llap{\raise.6ex\hbox{\smdag\char'171}}}
\def\tc{\llap{\raise.6ex\hbox{\smdag\char'172}}}
\def\td{\llap{\raise.6ex\hbox{\smdag\char'170}}}
\def\te{\llap{\raise.6ex\hbox{\smdag\char'173}}}
\def\tf{\llap{\raise.6ex\hbox{\smdag\char'153}}}

\def\ti{\llap{\raise.6ex\hbox{\smdag\char'172\char'172}}}
\def\t#1{\llap{\raise.7ex\hbox{\ssrm #1}}\hs0}
\def\tC{\llap{\raise.7ex\hbox{\ssrm C}}}
\def\tI{\llap{\raise.7ex\hbox{\ssrm I}}}
\def\tN{\llap{\raise.7ex\hbox{\ssrm N}}}
\def\tT{\llap{\raise.7ex\hbox{\ssrm T}}}
\def\tU{\llap{\raise.7ex\hbox{\ssrm U}}}
\def\tW{\llap{\raise.7ex\hbox{\ssrm W}}}
\def\thsp{\thinspace}    \def\msp{\kern.4em}     \def\sk{\hskip.6em\relax}
\def\kms{km~s$^{-1}$}    \def\kmss{km~s$^{-1}$\kern.25em}
\def\itkms{\hbox{{\it km s}\raise.6ex\hbox{\sit\hs{0.8}--\hs{0.5}1}}\hs1}

\def\etal{{\it et al\/}.\hs3}       \def\setal{{\sit et al\/}.}
\def\etal#1{{\it et al\/}.\hs{-1.3}$^{#1}$}  
\def\b-v{($B-V$)}                   \def\u-b{($U-B$)}
\def\sec{\hbox{$''$\hskip-3pt .}}   
    
             \def\Amms{\AA~mm$^{-1}$\kern .35em}
\def\msun{$M_{\odot}$\thsp}         \def\rsun{$R_{\odot}$\thsp}

\def\dego{\raise.9ex\hbox{\math\char'16}}
\def\hs#1{\hskip#1pt}               \def\hm#1{\hskip#1em} 
\def\frac#1/#2{\leavevmode\kern.05em\raise.6ex\hbox{\the\scriptfont0
    #1}\kern-.15em/\kern-.15em\lower.3ex\hbox{\the\scriptfont0 #2}}
\def\smfrac#1/#2{\leavevmode\kern.05em\raise.4ex\hbox{\the\scriptscriptfont0
    #1}\kern-.15em/\kern-.15em\lower.2ex\hbox{\the\scriptscriptfont0 #2}}

\def\cor{\hbox{{\it C\hs{-1} oravel\/}}}      \def\cors{\cor\hs4}
\def\smcor{\hbox{\sit C\hs{-0.5}oravel\/}}    \def\smcors{\smcor\hs4}
\def\m{\hbox{\raise.8ex\hbox{\srm m}}}
\def\mag{\hbox{\raise.8ex\hbox{\srm m}\hskip-1pt .}}
\def\x{\hbox{\raise.2ex\hbox{,,}}}

\hsize 16truecm\hoffset-0.6truecm\vsize26truecm\voffset-0.8truecm
\tabskip1.9em
\brm\cl{\hm2T{\rm ABLE} I}\ms\bit
\cl{\hm2Radial-velocity observations of HD 152028 \hs2 \brm (\hs{-1}\bit GK
Draconis\/\brm )                             }\bs\rm
\halign{#\hfil\hm{-2}&\hfil#&\hfil#&\hfil$#$\hm{-0.4}&\hfil$#$&\hfil$#$&
%          YYYY  MM      DD    MJD       V1               V2      Phi1
\hfil$#$&\hfil$#$&\hfil$#$\hm{-0.4}&\hfil$#$\cr
% Phi2   Vpuls    O-C1               O-C2
\it\hs{-1}Heliocentric Date\span& \it HMJD\hm{0.4}&Velocity\hm{0.3}\span&
\it Phase\span&V_{pulse}\hm{-0.3}& \rm (\hs1\it O\hs1-\hs2C\rm\hs1)\hs2\span\cr
&&&Prim.\hm{-0.25}&Sec.\hs2&Orbit&Pulse\hm{0.4} &       &Prim.\hm{-0.5}&Sec.\cr
&&&\itkms\hm{-0.4}&                              \itkms\hm{-0.4}&
  &                                     &         \itkms\hm{-0.6}
  &\itkms\hm{-0.6}&                              \itkms\hm{-0.8}\cr
%\noalign{\vs{-2}\rule\vs{-2}}
\noalign{\filbreak\vs{12}}
 2001\hs4May  &13.120&  52042.120&    -59.1&    +80.7&  0.240&   0.884&+2.2&
    +0.6&    -1.7\cr
\hm2 \hs4June &18.040&    078.040&    +67.5&    -76.4&  3.841& 305.871&+2.4&
    -0.1&    +0.0\cr
\hm2 \hs4     &23.034&    083.034&    -54.3&    +70.9&  4.342& 348.276&-2.0&
    -0.5&    +1.2\cr
\hm2 \hs4     &25.005&    085.005&      +2.0\hs4\span   &.540& 365.012&-0.7&
$---$\hm{0.5}&$---$\hm{0.5}\cr
\hm2 \hs4     &28.020&    088.020&    +65.6&    -73.4   &.842& 390.608&+1.2&
    -0.8&    +2.9\cr
\hm2 \hs4July &15.994&    105.994&    +41.0&    -43.0&  6.644& 543.224&-2.4&
    +1.3&    +4.6\cr
\hm2 \hs4     &25.971&    115.971&    +43.8&    -53.6&  7.645& 627.936&+1.4&
    +0.2&    -5.9\cr
\hm2 \hs4     &26.962&    116.962&    +63.2&    -68.4   &.744& 636.349&-1.3&
    +0.3&    +6.8\cr
\hm2 \hs4     &28.030&    118.030&    +64.0&    -77.2   &.851& 645.415&-0.7&
    +0.8&    -2.4\cr
\hm2 \hs4     &30.034&    120.034&    -12.5&    +13.8&  8.052& 662.429&-0.5&
    +0.8&    -7.1\cr
\hm2 \hs4Aug. & 1.042&    122.042&    -60.7&    +78.4   &.253& 679.481& 0.0&
    +1.2&    -4.0\cr
\hm2 \hs4     & 1.918&    122.918&    -49.4&    +68.9   &.341& 686.920&+1.7&
    +0.9&    -1.1\cr
\hm2 \hs4Nov. & 1.769&    214.769&    +13.2&    -13.2& 17.550&1466.805&+2.5&
    -0.8&    -4.0\cr
\noalign{\filbreak\vs{12}}
 2002\hs4Apr. &27.094&  52391.094&    -60.4&    +75.1& 35.229&2963.932&+1.5&
    -0.2&    -6.9\cr
\hm2 \hs4May  &28.070&    422.070&    -51.6&    +70.8& 38.335&3226.943&+1.2&
    +0.4&    -0.7\cr
\hm2 \hs4July &26.943&    481.943&    -54.8&    +73.0& 44.338&3735.309&-1.7&
    -0.5&    +2.3\cr
\hm2 \hs4Sept.& 2.900&    519.900&    -47.2&    +68.0& 48.144&4057.589&+1.1&
    -0.4&    +3.2\cr
\hm2 \hs4     & 3.949&    520.949&    -61.1&    +79.1   &.249&4066.498&+0.2&
    +0.7&    -3.4\cr
\hm2 \hs4     & 4.965&    521.965&    -52.6&    +68.3   &.351&4075.126&-2.6&
    +0.1&    +0.8\cr
\hm2 \hs4     &12.833&    529.833&    -45.9&    +63.1& 49.139&4141.928&+1.5&
    -0.8&    -0.2\cr
\hm2 \hs4     &26.892&    543.892&    +10.3&     -5.0& 50.549&4261.304&-1.8&
    +1.1&    +3.7\cr
\hm2 \hs4     &28.877&    545.877&    +61.5&    -73.5   &.748&4278.156&-2.7&
    -0.6&    +2.4\cr
\hm2 \hs4Oct. & 6.895&    553.895&     +9.5&     -7.1& 51.552&4346.237&-2.3&
    -0.2&    +2.9\cr
\hm2 \hs4     &12.869&    559.869&    -48.8&    +68.7& 52.151&4396.956&+0.9&
    +0.1&    +1.4\cr
\hm2 \hs4     &23.800&    570.800&    -59.4&    +83.5& 53.247&4489.770&+2.5&
    +0.2&    +1.0\cr
\hm2 \hs4Nov. & 4.784&    582.784&    -23.7&    +36.0& 54.448&4591.526&+0.5&
    -0.9&    +1.9\cr
\hm2 \hs4Dec. & 4.741&    612.741&    -20.1&    +35.2& 57.452&4845.883&+2.2&     -0.2&    +2.5\cr
\hm2 \hs4     & 9.720&    617.720&    +30.7&    -34.0   &.951&4888.153&-2.7&
    -0.8&    +3.6\cr
\noalign{\filbreak\vs{12}}
 2003\hs4Jan. &11.278&  52650.278&    -59.6&    +86.3& 61.216&5164.596&+1.1&
    0.0&    +5.4\cr
\hm2 \hs4     &16.250&    655.250&    +63.0&    -69.1   &.714&5206.814&+2.6&
    +1.1&     0.0\cr
\hm2 \hs4     &28.237&    667.237&    +49.6&    -53.4& 62.916&5308.589&+1.1&
    +0.9&    +1.1\cr
\hm2 \hs4Feb. &15.227&    685.227&    +57.0&    -73.4& 64.720&5461.341&-1.4&
    -2.0&    -3.0\cr
\hm2 \hs4     &19.089&    689.089&    -38.7&    +55.0& 65.107&5494.129&-2.6&
    -0.1&    +5.1\cr
\hm2 \hs4     &20.078&    690.078&   -59.8&$---$\hm{0.5}&.206&5502.528&+0.5&
    -0.4&$---$\hm{0.5}\cr
\hm2 \hs4     &21.206&    691.206&    -57.7&    +82.6   &.319&5512.107&-2.5&
    +0.7&    +7.7\cr
\hm2 \hs4     &22.089&    692.089&    -35.0&    +46.8   &.408&5519.606&+1.2&
    -0.5&    -2.8\cr
\hm2 \hs4Mar. & 1.223&    699.223&    -44.6&    +58.3& 66.123&5580.174&-2.7&
    -0.2&    +1.4\cr
\hm2 \hs4     & 3.196&    701.196&    -54.7&    +76.5   &.321&5596.930&+1.5&
    -0.5&    +2.0\cr
\hm2 \hs4     &15.208&    713.208&    +5.3\hs4\span&  67.525&5698.922&+1.6&
$---$\hm{0.5}&$---$\hm{0.5}\cr
\hm2 \hs4     &16.160&    714.160&    +34.5&    -35.3   &.621&5707.005&-0.5&
    +0.1&    +3.3\cr
\hm2 \hs4     &17.102&    715.102&    +58.8&    -69.7   &.715&5715.002&-0.4&
    -0.3&    -0.4\cr
\hm2 \hs4     &19.140&    717.140&    +44.2&    -54.1   &.919&5732.303&-1.8&
    -0.5&    -1.1\cr
\hm2 \hs4     &20.123&    718.123&     +4.6\hs4\span&  68.018&5740.649&+1.6&
$---$\hm{0.5}&$---$\hm{0.5}\cr
\hm2 \hs4     &23.142&    721.142&    -57.6&    +73.2   &.321&5766.286&-1.9&
     0.0&    -1.4\cr
\hm2 \hs4     &24.000&    722.000&   -35.7&$---$\hm{0.5}&.407&5773.572&+0.9&
    -0.6&$---$\hm{0.5}\cr
\hm2 \hs4Apr. & 1.045&    730.045&    -59.5&    +85.9& 69.213&5841.877&+2.3&
    -1.2&    +5.2\cr
\hm2 \hs4     & 7.153&    736.153&    +70.0&    -80.7   &.826&5893.739&+2.3&
    +0.9&    -2.2\cr
\hm2 \hs4     & 8.069&    737.069&    +47.6&    -50.8   &.917&5901.515&+0.4&
    +0.2&    +3.0\cr
\hm2 \hs4     &16.108&    745.108&    +63.6&    -71.8& 70.724&5969.775&+2.5&
    +0.0&    -0.5\cr
\hm2 \hs4     &17.031&    746.031&    +67.8&    -77.2   &.816&5977.611&+1.3&
    -0.9&    +2.1\cr
}

\vfe
\hsize 13truecm\hoffset1truecm
\topskip15pt

\tabskip1.9em
\brm\cl{\hm2T{\rm ABLE} III}\ms\bit
\cl{\hm2Radial-velocity observations of HDE 284195 \hs4{\rm (}\hs{-1}V1094 Tau{\rm )}                 }\bs\rm
%\vs{-6}\rule\ss
\halign{#\hfil\hm{-2}&\hfil#&\hfil#&\hfil    $#$\hm{-0.4}&\hfil$#$&
\hfil#&\hfil$#$\hm{-0.4}&\hfil$#$\cr
\it\hs{-1}Heliocentric Date\span&               \it HMJD\hm{0.4}&Velocity\hm{0.3}\span&
\it Phase&                                        \rm (\hs1\it O\hs1-\hs2C\rm\hs1)\hs2\span\cr
&&&Prim.\hm{-0.25}&Sec.\hs1&                     &Prim.\hm{-0.5}&Sec.\cr
&&&\itkms\hm{-0.4}&                              \itkms\hm{-0.6}&
  &\itkms\hm{-0.6}&                              \itkms\hm{-0.8}\cr
%\noalign{\vs{-2}\rule\vs{-2}}
\noalign{\filbreak\vs{12}}
 2002\hs4Sept.&28.096&  52545.096&    -44.9&    +57.8&     0.632&
     0.0&    -0.6\cr
\hm2 \hs4     &30.149&    547.149&      +3.7\hs4\span&.861&
$---$\hm{0.5}&$---$\hm{0.5}\cr
\hm2 \hs4Oct. & 4.136&    551.136&      +2.9\hs4\span&     1.304&
$---$\hm{0.5}&$---$\hm{0.5}\cr
\hm2 \hs4     &18.107&    565.107&      +3.4\hs4\span&     2.858&
$---$\hm{0.5}&$---$\hm{0.5}\cr
\hm2 \hs4     &19.097&    566.097&    +64.8&    -61.1&.969&
     0.0&    -0.2\cr
\hm2 \hs4     &24.085&    571.085&    -40.2&    +53.9&     3.524&
    +0.2&    +0.5\cr
\hm2 \hs4     &28.144&    575.144&    +68.0&    -64.4&.975&
    -0.2&    +0.1\cr
\hm2 \hs4Nov. & 2.127&    580.127&    -40.1&    +54.9&     4.530&
    +0.8&    +0.9\cr
\hm2 \hs4     & 7.070&    585.070&    +80.7&    -78.2&     5.080&
     0.0&     0.0\cr
\hm2 \hs4     &11.035&    589.035&    -40.5&    +53.1&.521&
    -0.4&    -0.1\cr
\hm2 \hs4Dec. & 4.974&    612.974&    +41.9&    -34.5&     8.184&
    -0.2&    +1.6\cr
\hm2 \hs4     & 9.883&    617.883&    -38.2&    +49.3&.730&
    -0.8&    -1.0\cr
\hm2 \hs4     &11.083&    619.083&      +5.1\hs4\span&.864&
$---$\hm{0.5}&$---$\hm{0.5}\cr
\noalign{\filbreak\vs{12}}
 2003\hs4Jan. & 5.000&  52644.000&    -45.3&    +58.4&    11.636&
    -0.4&     0.0\cr
\hm2 \hs4     & 5.978&    644.978&    -35.6&    +47.8&.745&
    -0.6&    +0.2\cr
\hm2 \hs4     & 7.000&    646.000&      +4.1\hs4\span&.858&
$---$\hm{0.5}&$---$\hm{0.5}\cr
\hm2 \hs4     & 7.853&    646.853&    +56.4&    -51.2&.953&
    +0.1&    +0.5\cr
\hm2 \hs4     &10.023&    649.023&    +37.6&    -31.3&    12.195&
    -0.1&    +0.2\cr
\hm2 \hs4     &10.975&    649.975&      +4.1\hs4\span&.301&
$---$\hm{0.5}&$---$\hm{0.5}\cr
\hm2 \hs4     &11.766&    650.766&    -21.8&    +32.0&.389&
    -1.2&    +0.1\cr
\hm2 \hs4     &11.890&    650.890&    -22.9&    +34.2&.403&
    +0.4&    -0.7\cr
\hm2 \hs4     &15.757&    654.757&     -9.4&    +21.0&.833&
    +0.5&    +0.6\cr
\hm2 \hs4     &16.776&    655.776&    +51.5&    -45.9&.946&
    -0.5&    +1.1\cr
\hm2 \hs4     &17.792&    656.792&    +84.4&    -83.3&    13.059&
    -0.1&    -1.0\cr
\hm2 \hs4     &20.846&    659.846&    -22.1&    +34.4&.399&
    +0.5&    +0.2\cr
\hm2 \hs4     &23.753&    662.753&    -38.3&    +52.7&.722&
    +0.3&    +1.2\cr
\hm2 \hs4     &25.760&    664.760&    +51.7&    -47.7&.946&
    -0.1&    -1.0\cr
\hm2 \hs4     &26.940&    665.940&    +81.4&    -77.6&    14.077&
    +0.1&    +1.2\cr
\hm2 \hs4     &27.836&    666.836&    +45.1&    -38.5&.177&
     0.0&    +0.9\cr
\hm2 \hs4Feb. & 1.816&    671.816&    -37.1&    +50.1&.731&
    +0.3&    -0.1\cr
\hm2 \hs4     &13.901&    683.901&    +82.0&    -79.3&    16.075&
    +0.3&     0.0\cr
\hm2 \hs4     &14.876&    684.876&    +43.3&    -35.1&.184&
    +1.1&    +1.2\cr
\hm2 \hs4     &17.838&    687.838&    -39.3&    +52.1&.513&
    +0.1&    -0.3\cr
\hm2 \hs4     &18.834&    688.834&    -45.0&    +59.9&.624&
     0.0&    +1.4\cr
\hm2 \hs4     &19.798&    689.798&    -36.9&    +49.4&.731&
    +0.4&    -0.7\cr
\hm2 \hs4     &20.859&    690.859&     -2.3&    +13.1&.849&
    +0.5&    +0.5\cr
\hm2 \hs4     &21.870&    691.870&    +60.9&    -58.9&.962&
    -0.3&    -2.0\cr
\hm2 \hs4     &22.840&    692.840&    +82.7&    -80.6&    17.070&
    -0.1&    -0.1\cr
\hm2 \hs4Mar. &14.815&    712.815&      +4.4\hs4\span&    19.292&
$---$\hm{0.5}&$---$\hm{0.5}\cr
\hm2 \hs4     &15.815&    713.815&    -23.3&    +34.3&.403&
    +0.2&    -0.8\cr
\hm2 \hs4     &16.819&    714.819&    -39.8&    +52.7&.515&
    -0.2&    +0.1\cr
\hm2 \hs4     &19.823&    717.823&     -3.8&    +10.9&.849&
    -1.0&    -1.7\cr
\hm2 \hs4     &23.816&    721.816&      +4.3\hs4\span&    20.294&
$---$\hm{0.5}&$---$\hm{0.5}\cr
\hm2 \hs4Apr. & 2.831&    731.831&    -24.4&    +36.5&    21.408&
    -0.1&    +0.5\cr
}

\bye
\bye
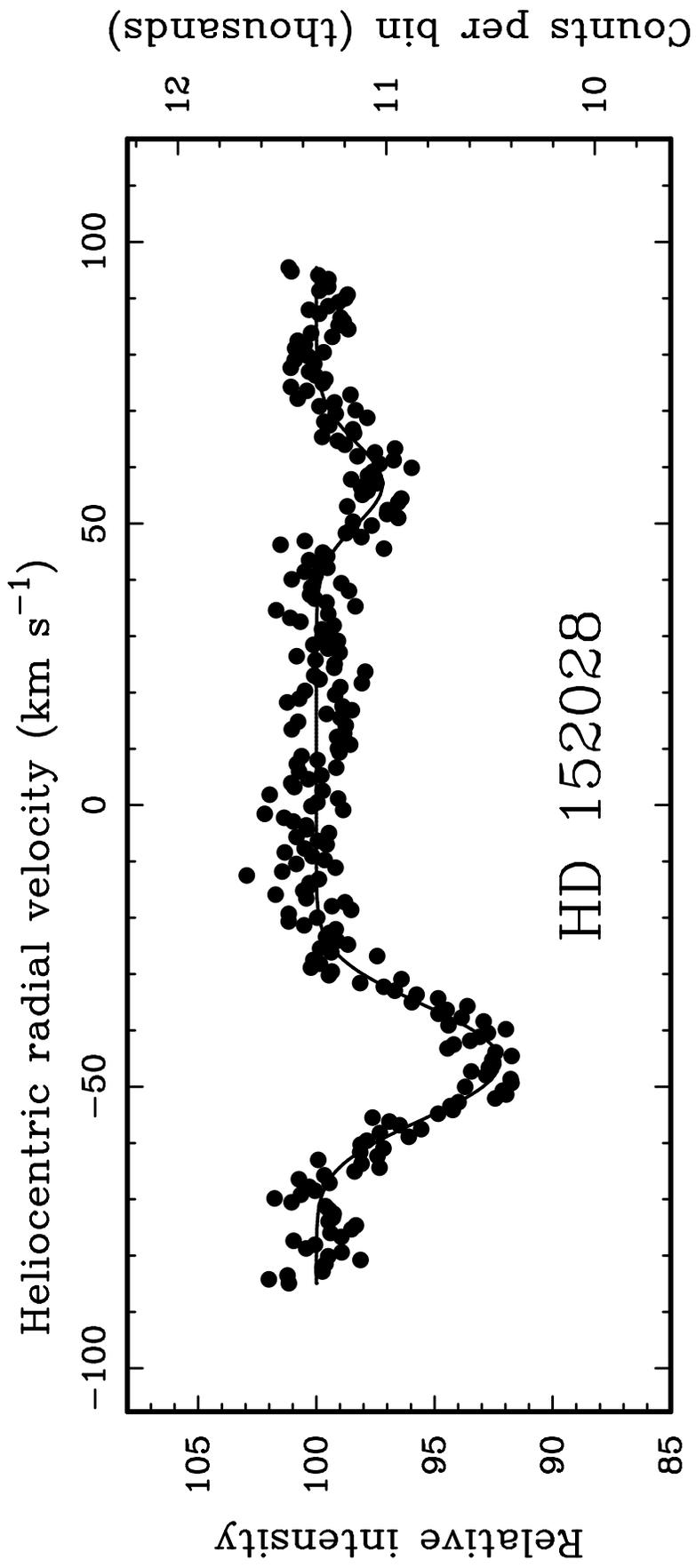
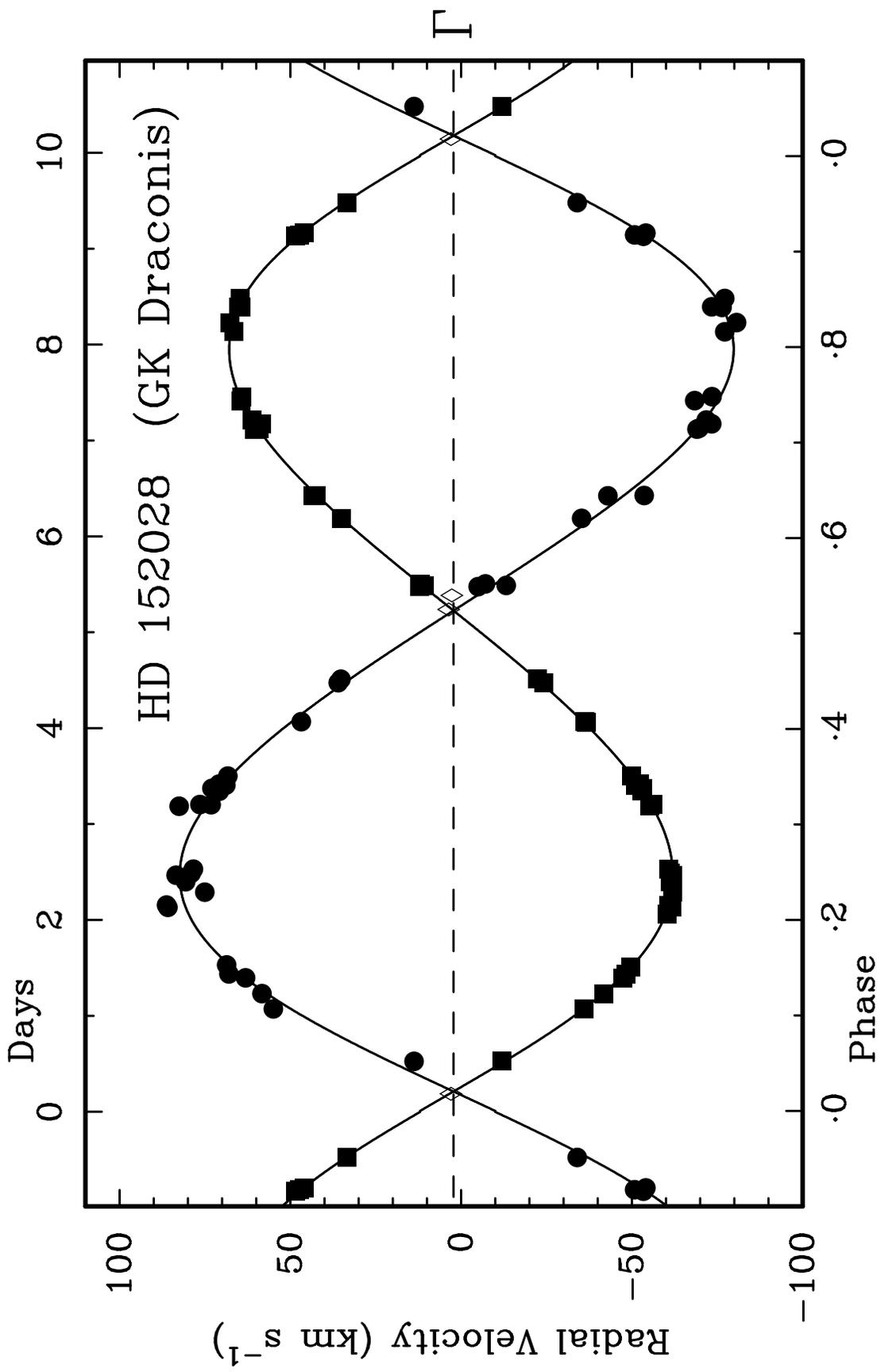
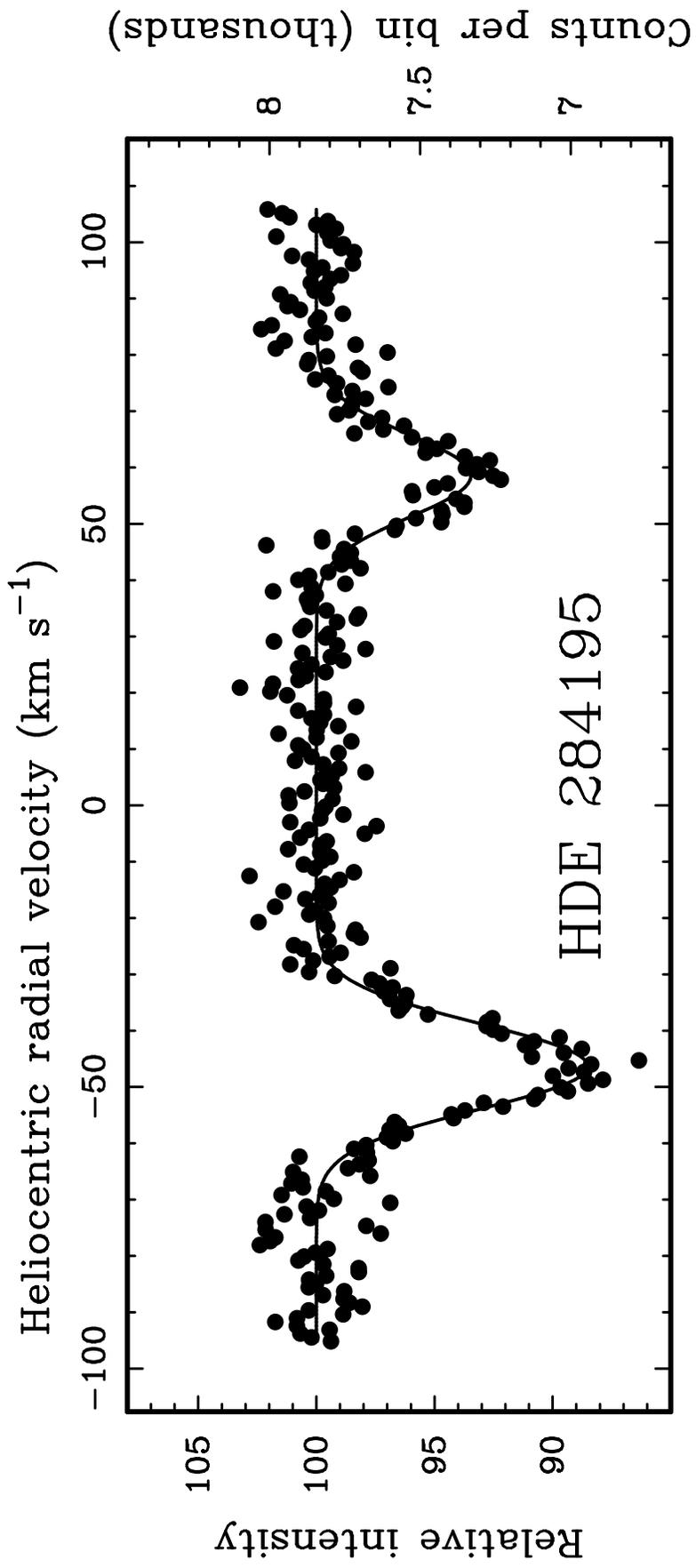
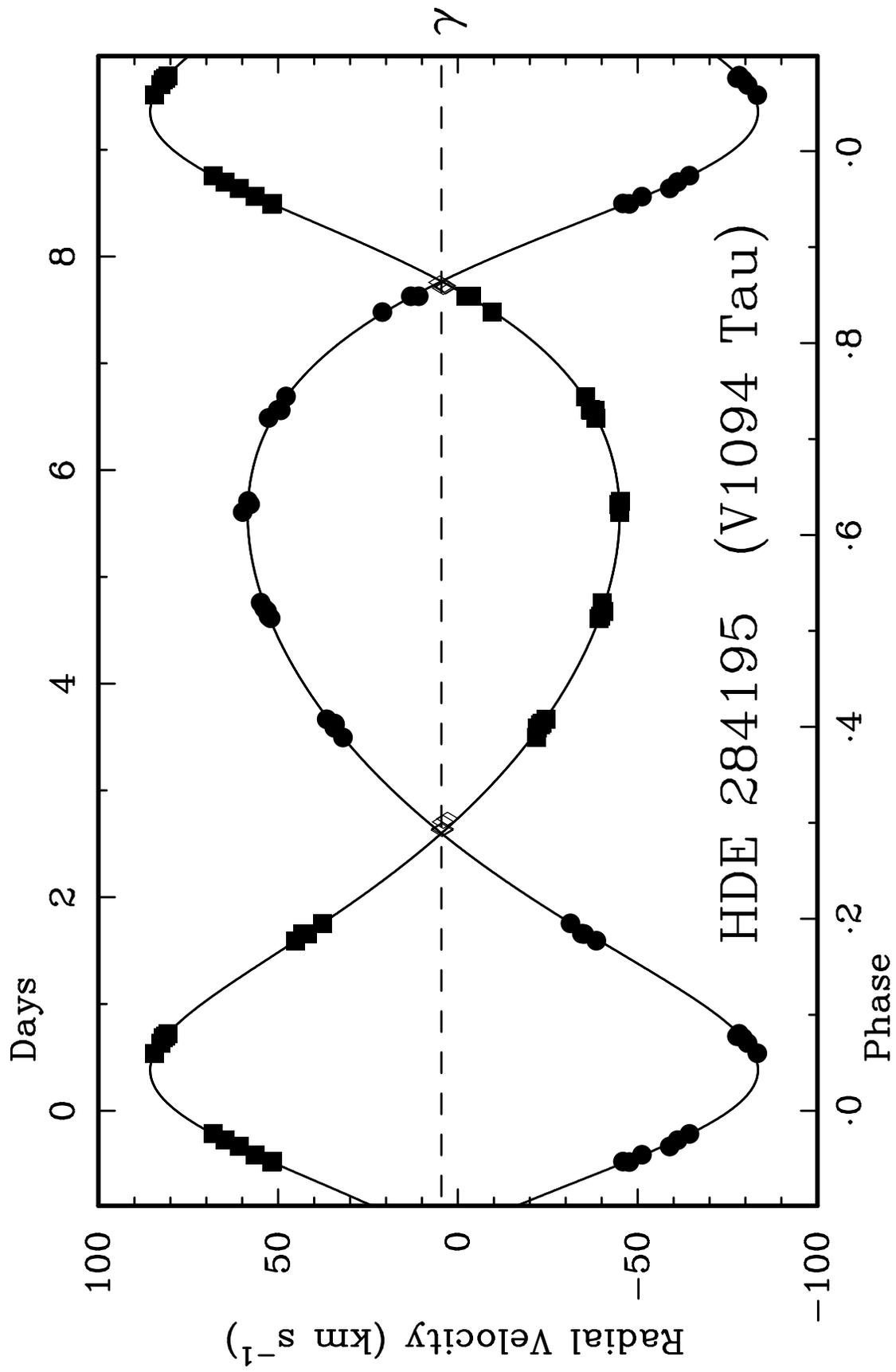
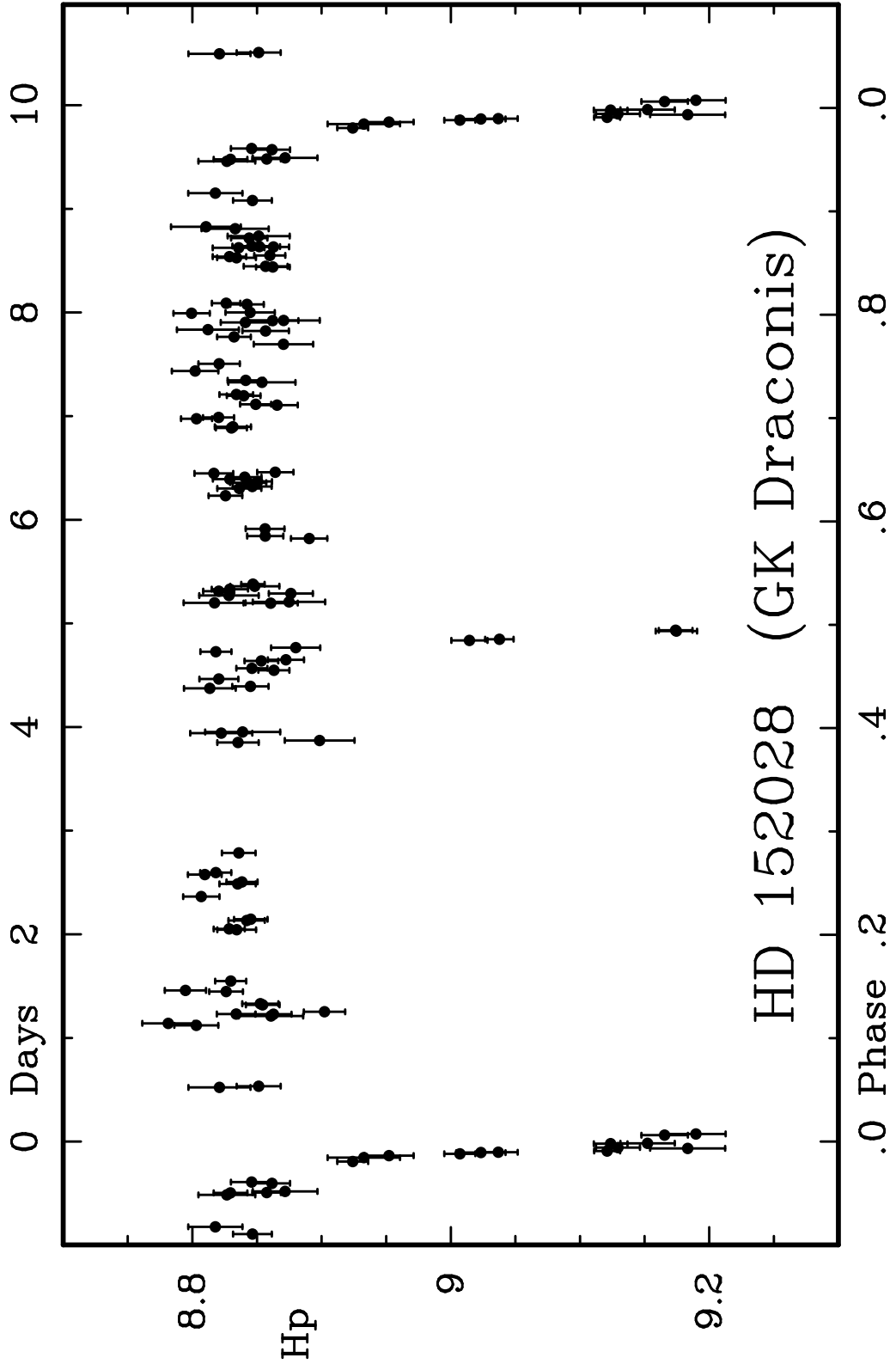
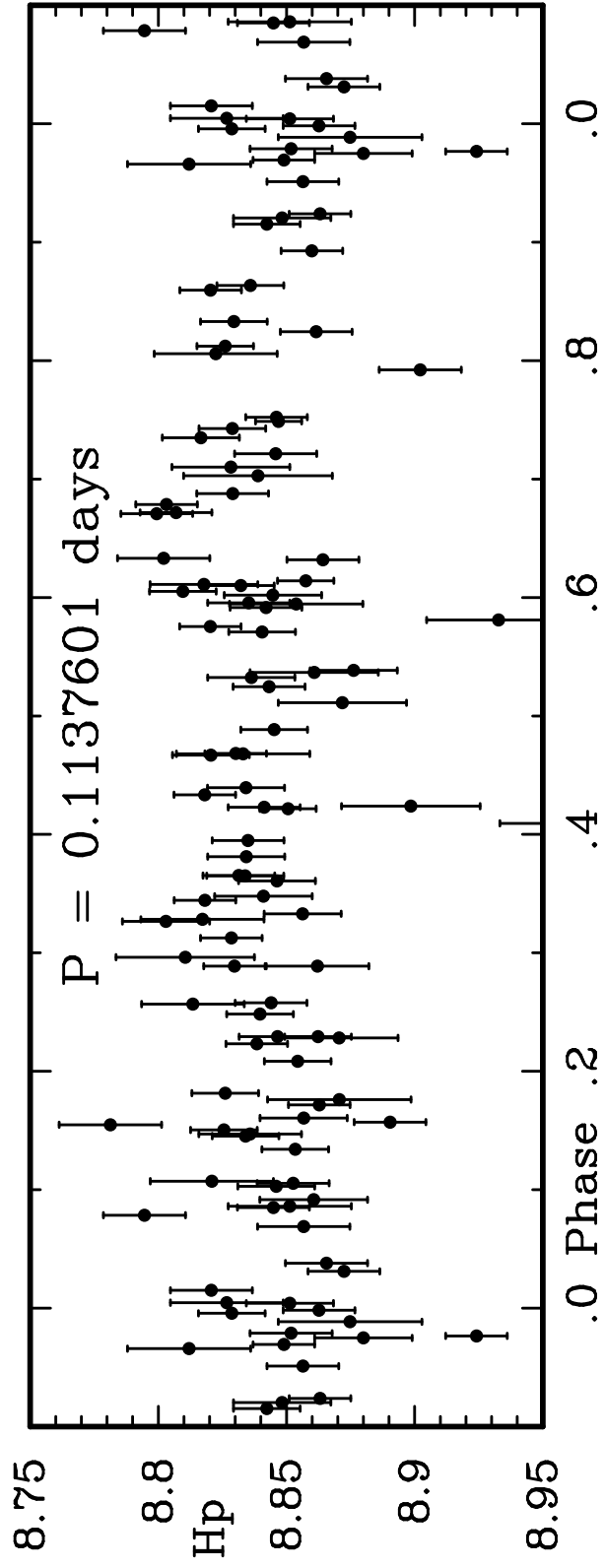
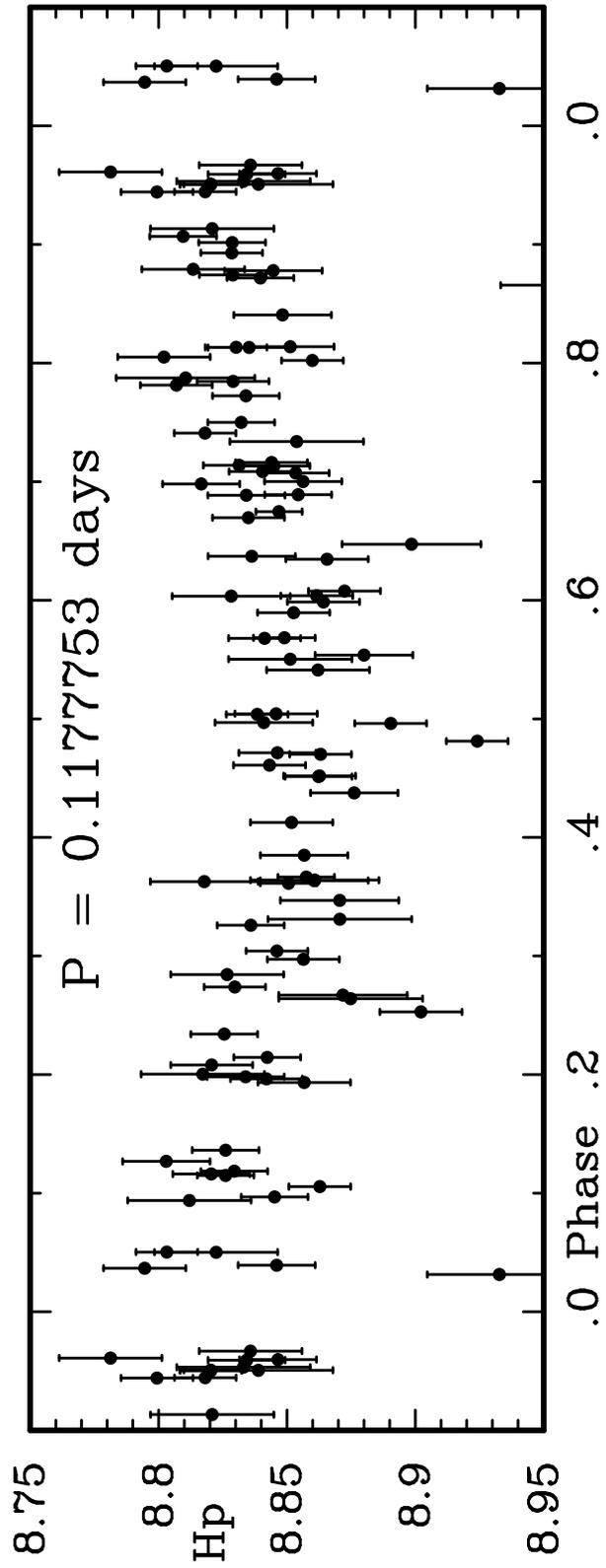
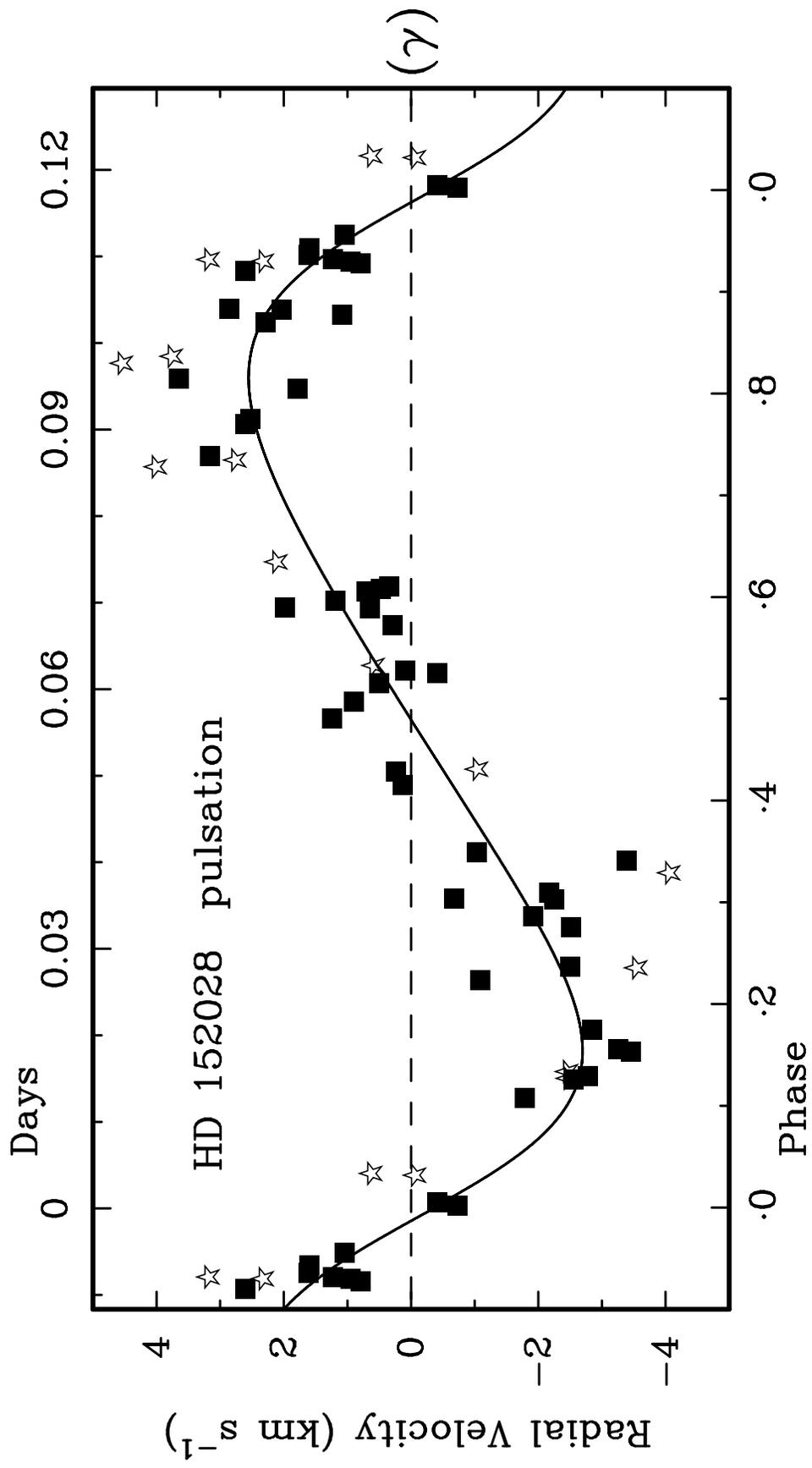
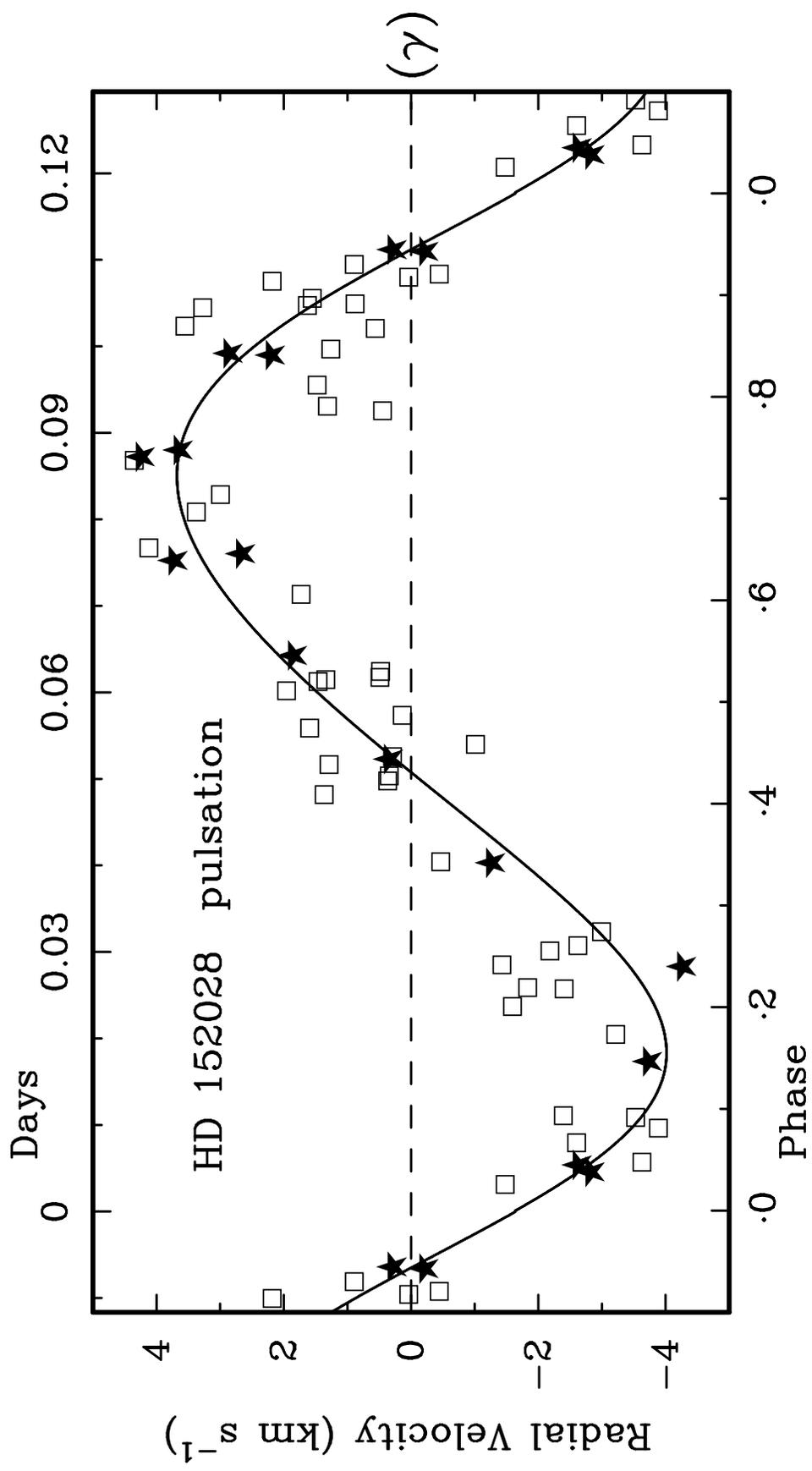
\end